\documentclass[numberedappendix]{aastex61}
\usepackage{color}
\usepackage{times}
\usepackage{xspace}
\usepackage{amsmath}
\submitjournal{ApJ}

\shorttitle{Lensed $z=2.22$ SN~Ia behind MOO~J1014+0038}
\shortauthors{Rubin and Hayden et al.}

\begin{document}

\newcommand{\ang}{\AA\xspace}
\newcommand{\HST}{\textit{HST}\xspace}
\newcommand{\Hubble}{\textit{Hubble Space Telescope}\xspace}
\newcommand{\redshift}{2.2216\xspace}
\newcommand{\shortredshift}{2.22\xspace}
\newcommand{\initmagnification}{$\sim 2.2$\xspace}
\newcommand{\finaldistmod}{$45.11 \pm 0.22$ mag\xspace}
\newcommand{\finalmagnification}{$1.10 \pm 0.23$ mag\xspace}
\newcommand{\finalmagnificationflux}{$2.8^{+0.6}_{-0.5}$\xspace}
\newcommand{\fIa}{\ensuremath{f_{\mathrm{\, Ia}}}\xspace}
\newcommand{\clusterLL}{$>100$\xspace}
\newcommand{\supernovaBGCoffset}{15\arcsec\xspace}
\newcommand{\sersic}{S\'{e}rsic\xspace}
\newcommand{\weaklensingonlyamplflux}{1.75\xspace}
\newcommand{\fastage}{\ensuremath{1.60 \pm 0.1}\xspace}
\newcommand{\fasttau}{\ensuremath{0.16 \pm 0.02}\xspace}
\newcommand{\fastAv}{\ensuremath{0.3 \pm 0.08}\xspace}
\newcommand{\fastAvthree}{\ensuremath{1.0 \pm 0.27}\xspace}
\newcommand{\fastmass}{\ensuremath{11.08 \pm 0.02}\xspace}
\newcommand{\ppxfvdisp}{\ensuremath{305 \pm 42}\xspace}
\newcommand{\ppxfvdispnoerror}{305 km/s\xspace}
\newcommand{\nsersic}{\ensuremath{2.8 \pm 0.1}\xspace}
\newcommand{\halflight}{\ensuremath{0 \farcs 225 \pm 0.006}\xspace}
\newcommand{\halflightdemag}{\ensuremath{0.679 \pm 0.017}\xspace}
\newcommand{\centraldensity}{\ensuremath{10.62 \pm 0.02}\xspace}
\newcommand{\type}{\mathrm{type}}
\newcommand{\IaAll}{\mathrm{All\, Ia}}
\newcommand{\Ia}{\mathrm{Ia}}
\newcommand{\IaT}{\mathrm{Ia\ 91T}}
\newcommand{\Ibc}{\mathrm{Ib/c}}
\newcommand{\II}{\mathrm{II}}
\newcommand{\Ltype}{\ensuremath{\overline{L}_{\type}}\xspace}
\newcommand{\LtypeIa}{\ensuremath{\overline{L}_{\mathrm{Ia}}\xspace}}
\newcommand{\Ltypebg}{\ensuremath{\overline{L}_{\mathrm{Ia\ 91bg}}\xspace}}
\newcommand{\LtypeT}{\ensuremath{\overline{L}_{\mathrm{Ia\ 91T}}\xspace}}
\newcommand{\LtypeIbc}{\ensuremath{\overline{L}_{\mathrm{Ibc}}\xspace}}
\newcommand{\LtypeII}{\ensuremath{\overline{L}_{\mathrm{II}}\xspace}}

\newcommand{\SSFR}{\ensuremath{-11.48^{+0.17}_{-0.30}}\xspace}
\newcommand{\oiimean}{\ensuremath{1.5^{+3.0}_{-2.8}}\xspace}
\newcommand{\oiilum}{\ensuremath{3.3^{+7.6}_{-6.2} \times 10^{41} }\xspace}
\newcommand{\halphamean}{\ensuremath{1.1^{+1.9}_{-1.7}}\xspace}
\newcommand{\halphalum}{\ensuremath{2.4^{+4.1}_{-3.7} \times 10^{41} }\xspace}
\newcommand{\FASTSFR}{\ensuremath{0.4\pm 0.2}\xspace}
\newcommand{\HalphaSFR}{\ensuremath{1.3^{+2.3}_{-2.0}}\xspace}
\newcommand{\OIISFR}{\ensuremath{9^{+20}_{-17}}\xspace}
\newcommand{\CombinedSFR}{\ensuremath{0.4\pm 0.2}\xspace}
\newcommand{\halphaSEDagree}{\ensuremath{0.4}\xspace}
\newcommand{\LmeanIa}{1\xspace}
\newcommand{\LmeanIaT}{2.67\xspace}
\newcommand{\LmeanIabg}{0\xspace}
\newcommand{\LmeanIbc}{0.05\xspace}
\newcommand{\LmeanII}{0.07\xspace}
\newcommand{\DzIa}{0.35\xspace}
\newcommand{\DzIaT}{0.23\xspace}
\newcommand{\DzIbc}{1.93\xspace}
\newcommand{\DzII}{0.93\xspace}
\newcommand{\hostIaRateRest}{0.0065\xspace}
\newcommand{\hostCCRateRest}{0.0028\xspace}
\newcommand{\hostArcSecIaRateRest}{0.00065\xspace}
\newcommand{\hostArcSecCCRateRest}{0.00028\xspace}
\newcommand{\IaPercentHost}{68\%\xspace}
\newcommand{\IaTPercentHost}{2\%\xspace}
\newcommand{\IaCombinedPercentHost}{70\%\xspace}
\newcommand{\IaCombinedFractionHost}{\ensuremath{0.70}\xspace}
\newcommand{\IbcPercentHost}{7\%\xspace}
\newcommand{\IIPercentHost}{22\%\xspace}
\newcommand{\IaRateLocalHost}{0.00063\xspace}
\newcommand{\IaTRateLocalHost}{2$\times 10^{-5}$\xspace}
\newcommand{\IbcRateLocalHost}{7$\times 10^{-5}$\xspace}
\newcommand{\IIRateLocalHost}{0.00021\xspace}
\newcommand{\volumeArcsecIaRateRestNaive}{4.6$\times 10^{-5}$\xspace}
\newcommand{\volumeArcsecIaRateRestDelensed}{2.6$\times 10^{-5}$\xspace}
\newcommand{\volumeArcsecCCRateRestNaive}{0.00032\xspace}
\newcommand{\volumeArcsecCCRateRestDelensed}{0.00019\xspace}
\newcommand{\IaRateLocalProj}{8.9$\times 10^{-6}$ $\times$ 0.07\xspace}
\newcommand{\IaTRateLocalProj}{1.8$\times 10^{-7}$ $\times$ 0.07\xspace}
\newcommand{\IbcRateLocalProj}{8.9$\times 10^{-5}$ $\times$ 0.07\xspace}
\newcommand{\IIRateLocalProj}{0.00013 $\times$ 0.07\xspace}
\newcommand{\IaRateHostUnnorm}{0.00063\xspace}
\newcommand{\IaTRateHostUnnorm}{5.2$\times 10^{-5}$\xspace}
\newcommand{\IbcRateHostUnnorm}{3.7$\times 10^{-6}$\xspace}
\newcommand{\IIRateHostUnnorm}{1.4$\times 10^{-5}$\xspace}
\newcommand{\IaRateProjUnnorm}{6.2$\times 10^{-7}$\xspace}
\newcommand{\IaTRateProjUnnorm}{3.4$\times 10^{-8}$\xspace}
\newcommand{\IbcRateProjUnnorm}{3.3$\times 10^{-7}$\xspace}
\newcommand{\IIRateProjUnnorm}{6$\times 10^{-7}$\xspace}
\newcommand{\IaRateHostNorm}{0.898\xspace}
\newcommand{\IaTRateHostNorm}{0.074\xspace}
\newcommand{\IbcRateHostNorm}{0.005\xspace}
\newcommand{\IIRateHostNorm}{0.020\xspace}
\newcommand{\finalIaAllprob}{97.5\%\xspace}
\newcommand{\finalCCAllprob}{2.5\%\xspace}
\newcommand{\finalIaNprob}{90.1\%\xspace}
\newcommand{\finalIaTprob}{7.4\%\xspace}
\newcommand{\finalIbcprob}{0.5\%\xspace}
\newcommand{\finalIIprob}{2.0\%\xspace}
\newcommand{\IaRateProjNorm}{0.001\xspace}
\newcommand{\IaTRateProjNorm}{0.000\xspace}
\newcommand{\IbcRateProjNorm}{0.000\xspace}
\newcommand{\IIRateProjNorm}{0.001\xspace}

\title{The Discovery of a Gravitationally Lensed Supernova Ia at Redshift \shortredshift}

\correspondingauthor{David Rubin}
\email{drubin@stsci.edu}

\author{D. Rubin}
\affiliation{D. Rubin and B. Hayden were equal coauthors.}
\affiliation{Space Telescope Science Institute, 3700 San Martin Drive, Baltimore, MD 21218}
\affiliation{E.O. Lawrence Berkeley National Lab, 1 Cyclotron Rd., Berkeley, CA, 94720}
\author{B. Hayden}
\affiliation{D. Rubin and B. Hayden were equal coauthors.}
\affiliation{E.O. Lawrence Berkeley National Lab, 1 Cyclotron Rd., Berkeley, CA, 94720}
\affiliation{Department of Physics, University of California Berkeley, Berkeley, CA 94720}
\author{X. Huang}
\affiliation{Department of Physics and Astronomy, University of San Francisco, San Francisco, CA 94117-1080}
\author{G. Aldering}
\affiliation{E.O. Lawrence Berkeley National Lab, 1 Cyclotron Rd., Berkeley, CA, 94720}
\author{R. Amanullah}
\affiliation{The Oskar Klein Centre, Department of Physics, AlbaNova, Stockholm University, SE-106 91 Stockholm, Sweden}
\author{K. Barbary}
\affiliation{E.O. Lawrence Berkeley National Lab, 1 Cyclotron Rd., Berkeley, CA, 94720}
\author{K. Boone}
\affiliation{E.O. Lawrence Berkeley National Lab, 1 Cyclotron Rd., Berkeley, CA, 94720}
\affiliation{Department of Physics, University of California Berkeley, Berkeley, CA 94720}
\author{M. Brodwin}
\affiliation{Department of Physics and Astronomy, University of Missouri-Kansas City, Kansas City, MO 64110}
\author{S. E. Deustua}
\affiliation{Space Telescope Science Institute, 3700 San Martin Drive, Baltimore, MD 21218}
\author{S. Dixon}
\affiliation{E.O. Lawrence Berkeley National Lab, 1 Cyclotron Rd., Berkeley, CA, 94720}
\affiliation{Department of Physics, University of California Berkeley, Berkeley, CA 94720}
\author{P. Eisenhardt}
\affiliation{Jet Propulsion Laboratory, California Institute of Technology, Pasadena, CA, 91109}
\author{A. S. Fruchter}
\affiliation{Space Telescope Science Institute, 3700 San Martin Drive, Baltimore, MD 21218}
\author{A. H. Gonzalez}
\affiliation{Department of Astronomy, University of Florida, Gainesville, FL 32611}
\author{A. Goobar}
\affiliation{The Oskar Klein Centre, Department of Physics, AlbaNova, Stockholm University, SE-106 91 Stockholm, Sweden}
\author{R. R. Gupta}
\affiliation{E.O. Lawrence Berkeley National Lab, 1 Cyclotron Rd., Berkeley, CA, 94720}
\author{I. Hook}
\affiliation{Physics Department, Lancaster University, Lancaster LA1 4YB, United Kingdom}
\author{M. J. Jee}
\affiliation{Department of Astronomy and Center for Galaxy Evolution Research, Yonsei University, 50 Yonsei-ro, Seoul 03772, Korea}
\author{A. G. Kim}
\affiliation{E.O. Lawrence Berkeley National Lab, 1 Cyclotron Rd., Berkeley, CA, 94720}
\author{M. Kowalski}
\affiliation{Institut f\"ur Physik, Newtonstr. 15, 12489 Berlin, Humboldt-Universit\"at zu Berlin, Germany}
\author{C. E. Lidman}
\affiliation{Australian Astronomical Observatory, PO Box 296, Epping, NSW 1710, Australia}
\author{E. Linder}
\affiliation{E.O. Lawrence Berkeley National Lab, 1 Cyclotron Rd., Berkeley, CA, 94720}
\author{K. Luther}
\affiliation{E.O. Lawrence Berkeley National Lab, 1 Cyclotron Rd., Berkeley, CA, 94720}
\affiliation{Department of Physics, University of California Berkeley, Berkeley, CA 94720}
\author{J. Nordin}
\affiliation{Institut f\"ur Physik, Newtonstr. 15, 12489 Berlin, Humboldt-Universit\"at zu Berlin, Germany}
\author{R. Pain}
\affiliation{Laboratoire de Physique Nucleaire et de Hautes Energies, Universite Pierre et Marie Curie, 75252 Paris Cedec 05, France}
\author{S. Perlmutter}
\affiliation{E.O. Lawrence Berkeley National Lab, 1 Cyclotron Rd., Berkeley, CA, 94720}
\affiliation{Department of Physics, University of California Berkeley, Berkeley, CA 94720}
\author{Z. Raha}
\affiliation{Department of Physics and Astronomy, University of San Francisco, San Francisco, CA 94117-1080}
\affiliation{E.O. Lawrence Berkeley National Lab, 1 Cyclotron Rd., Berkeley, CA, 94720}
\author{M. Rigault}
\affiliation{Institut f\"ur Physik, Newtonstr. 15, 12489 Berlin, Humboldt-Universit\"at zu Berlin, Germany}
\author{P.  Ruiz-Lapuente}
\affiliation{Institute of Cosmos Sciences, University of Barcelona, E-08028 Barcelona, Spain}
\affiliation{Instituto de Fisica Fundamental, CSIC, E-28006 Madrid, Spain}
\author{C. M. Saunders}
\affiliation{E.O. Lawrence Berkeley National Lab, 1 Cyclotron Rd., Berkeley, CA, 94720}
\affiliation{Department of Physics, University of California Berkeley, Berkeley, CA 94720}
\affiliation{Laboratoire de Physique Nucleaire et de Hautes Energies, Universite Pierre et Marie Curie, 75252 Paris Cedec 05, France}
\author{C. Sofiatti}
\affiliation{E.O. Lawrence Berkeley National Lab, 1 Cyclotron Rd., Berkeley, CA, 94720}
\affiliation{Department of Physics, University of California Berkeley, Berkeley, CA 94720}
\author{A. L. Spadafora}
\affiliation{E.O. Lawrence Berkeley National Lab, 1 Cyclotron Rd., Berkeley, CA, 94720}
\author{S. A. Stanford}
\affiliation{Department of Physics, University of California Davis, One Shields Avenue, Davis, CA 95616}
\author{D. Stern}
\affiliation{Jet Propulsion Laboratory, California Institute of Technology, Pasadena, CA, 91109}
\author{N. Suzuki}
\affiliation{Kavli Institute for the Physics and Mathematics of the Universe, University of Tokyo, Kashiwa, 277-8583, Japan}
\author{S. C. Williams}
\affiliation{Physics Department, Lancaster University, Lancaster LA1 4YB, United Kingdom}
\collaboration{(The Supernova Cosmology Project)}

\begin{abstract}

We present the discovery and measurements of a gravitationally lensed supernova (SN) behind the galaxy cluster MOO~J1014+0038. Based on multi-band \Hubble and Very Large Telescope (VLT) photometry of the supernova, and VLT spectroscopy of the host galaxy, we find a \finalIaAllprob probability that this SN is a SN~Ia, and a \finalCCAllprob chance of a CC SN. 
Our typing algorithm combines the shape and color of the light curve with the expected rates of each SN type in the host galaxy.
With a redshift of \redshift, this is the highest redshift SN~Ia discovered with a spectroscopic host-galaxy redshift. A further distinguishing feature is that the lensing cluster, at redshift 1.23, is the most distant to date to have an amplified SN. The SN lies in the middle of the color and light-curve shape distributions found at lower redshift, disfavoring strong evolution to $z=\shortredshift$. We estimate an amplification due to gravitational lensing of \finalmagnificationflux (\finalmagnification)---compatible with the value estimated from the weak-lensing-derived mass and the mass-concentration relation from $\Lambda$CDM simulations---making it the most amplified SN~Ia discovered behind a galaxy cluster.

\end{abstract}

\section{Introduction}

Gravitational lensing by massive galaxy clusters offers an amplified and magnified view of the high-redshift universe. Several examples of supernovae (SNe) lensed by foreground clusters have been found in recent years \citep{goobar09, amanullah11, nordin14, patel14, kelly15, rodney15, petrushevska16}. Four of these cluster-lensed SNe have been of Type~Ia (SNe~Ia), from which the amplification due to lensing has been determined. (Two additional SNe~Ia that were lensed by field galaxies have also been found, \citealt{quimby14, goobar16}; we summarize all referenced SNe in Table~\ref{tab:allsne}). 

The uniquely precise standardization possible with SNe~Ia provides amplification information, breaking the so-called mass-sheet degeneracy that is problematic for most shear-based lensing models, and permits independent direct tests of cluster mass models obtained from galaxy lensing \citep{nordin14, patel14, rodney15}. Lenses that produce multiple SN images provide additional model constraints \citep{kelly16}. To date the redshifts of gravitationally lensed background SNe~Ia have been at $z < 1.39$, a redshift range where even without lensing the complete (normal) SN~Ia population can be detected using single-orbit \HST visits. In cases of stronger amplification of sources beyond the redshift reach of normal observations, tests of SN~Ia properties and rates over a larger look-back time become possible. Here we report the discovery of the most amplified  SN~Ia behind a galaxy cluster ever found, with a spectroscopic host-galaxy redshift making it also the most distant.

\begin{deluxetable}{ccccccccc}
\rotate
\caption{Other high-redshift or gravitationally lensed SNe.}\label{tab:allsne}
\tablehead{\colhead{SN} & \colhead{Redshift} & \colhead{Redshift Type} & \colhead{Lens} & \colhead{Lens Redshift} & \colhead{Amplification} & \colhead{Type} & \colhead{Reference}}
\startdata
CAND-ISAAC & 0.64 & Spec & Cluster & 0.18 & 3.6 & IIP & \citet{goobar09, Stanishev09}  \\ 
PS1-10afx & 1.39 & Spec & Field & 1.12 & 30\tablenotemark{a} & Ia & \citet{quimby14} \\ 
A1, CLA11Tib & 1.14 & Spec & Cluster & 0.19 & 1.4 & Ia & \citet{nordin14, patel14} \\
H1, CLN12Did & 0.85 & Spec & Cluster & 0.35 & 1.3 & Ia & \citet{nordin14, patel14} \\
L2, CLO12Car & 1.28 & Spec & Cluster & 0.39 & 1.7 & Ia & \citet{nordin14, patel14} \\
Refsdal & 1.49 & Spec & Cluster & 0.54 & $\sim 100$\tablenotemark{a} & II & \citet{kelly15,kelly16b} \\ 
HFF14Tom & 1.35 & Spec & Cluster & 0.31 & 1.7 & Ia & \citet{rodney15} \\ 
GND13Sto &  $1.80 \pm 0.02$ & Phot & \nodata & \nodata & \nodata & Ia & \citet{rodney15b} \\ 
GND12Col &  $2.26^{+0.02}_{-0.10}$ & Phot & \nodata & \nodata &  \nodata & Ia & \citet{rodney15b} \\ 
CAND-669 & 0.67 & Spec & Cluster & 0.18 & 1.3 & IIL & \citet{petrushevska16}  \\ 
CAND-821 & 1.70 & Spec & Cluster & 0.18 & 4.3 & IIn & \citet{amanullah11, petrushevska16}  \\ 
CAND-1392 & 0.94 & Spec & Cluster & 0.18 & 2.7 & IIP & \citet{petrushevska16}  \\ 
CAND-10658 & $0.94^{+0.07}_{-0.27}$ & Spec & Cluster & 0.18 & 1.7 & IIn & \citet{petrushevska16}  \\ 
CAND-10662 & $1.03^{+0.20}_{-0.17}$ & Spec & Cluster & 0.18 & 2.6 & IIP & \citet{petrushevska16}  \\ 
iPTF16geu & 0.41 & Spec & Field & 0.22 & 52\tablenotemark{a} & Ia & \citet{goobar16} \\ 
SCP16C03 (``Joseph'') & \shortredshift & Spec & Cluster & 1.23 & 2.8 & Ia & This Work  \\
\enddata
\tablenotetext{a}{Multiple images; amplification is sum over all images.}
\end{deluxetable}

\section{Discovery}\label{sec:discovery}

The Supernova Cosmology Project's ``See~Change'' program (PI: Perlmutter) monitored twelve massive galaxy clusters in the redshift range 1.13 to 1.75 using the \Hubble (\HST) Wide Field Camera 3 (WFC3) with the goals of greatly expanding the range of the high-redshift SN~Ia Hubble diagram and accurately determining high-redshift cluster masses via weak lensing.  We observed each cluster every 5 weeks for one orbit split between the UVIS $F814W$, the IR $F105W$, and the IR $F140W$ filters.  The supernova survey was very deep; our 50\% completeness for simulated SN detection was AB 26.6 in $F105W + F140W$ \citep{hayden16}; this is $\sim 1$ magnitude fainter than a $z=1.75$ SN~Ia at maximum.\footnote{When stacking all the epochs together, we reach $5\sigma$ point-source depths of 28.0~AB~mag in $F105W$ and 27.9~AB~mag in $F140W$.} When a promising Type~Ia supernova candidate was found, at least one extra visit was triggered and executed within 2--3 observer frame weeks after the initial discovery. These extra visits provided better light-curve sampling, usually close to maximum light, and frequently extended the wavelength range with the IR $F160W$ filter.
 
\begin{figure*}[tbp]
\begin{centering}
\includegraphics[width= 0.9 \textwidth]{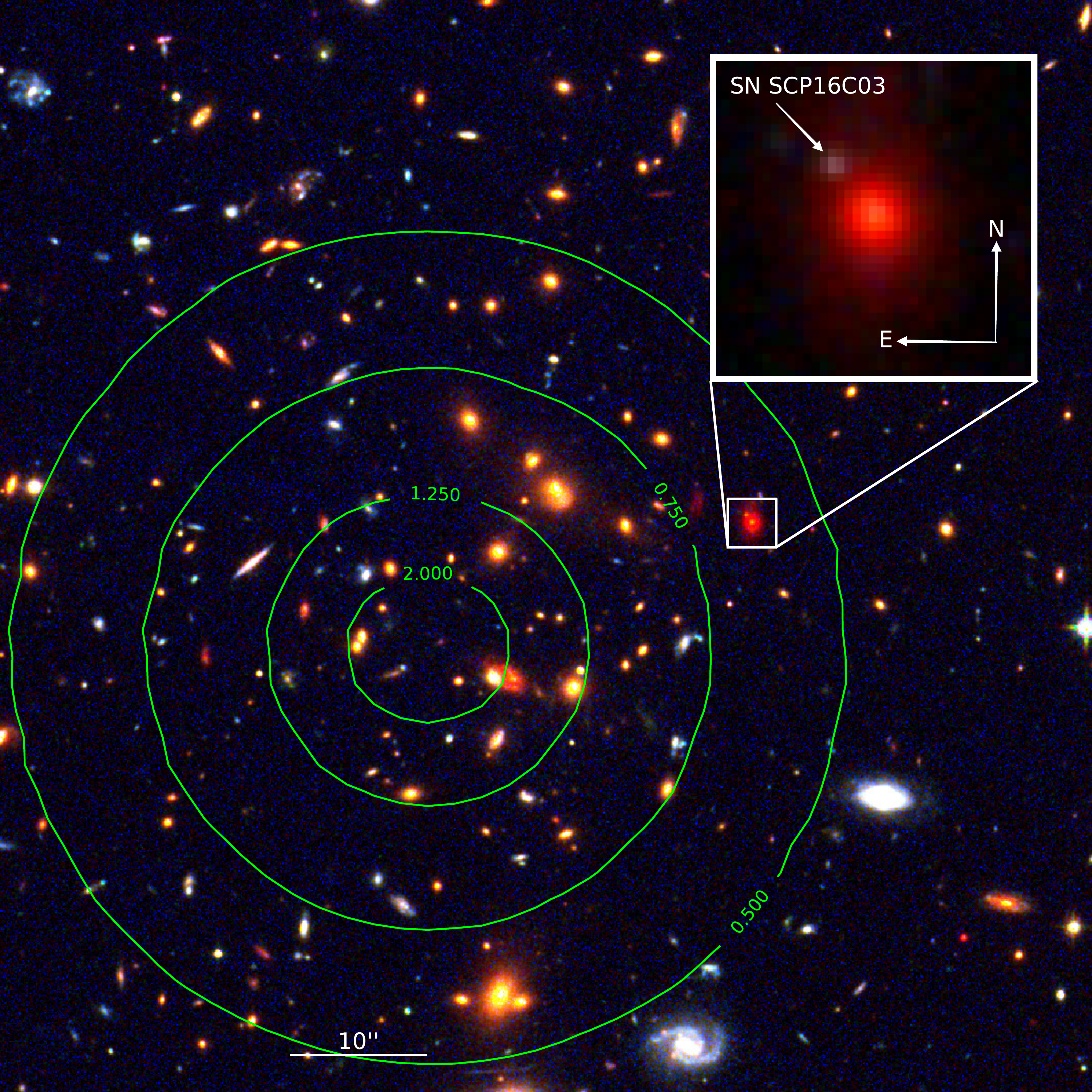}
\caption{Color image of the central part of MOO~J1014+0038 made with an $F814W$ stack and the $F105W$ and $F160W$ data from our first triggered \HST imaging near maximum light. The inset panel uses different scaling, and incorporates only IR data ($F105W$, $F125W$, and $F160W$). The compass arrows are 1\arcsec\ in linear scale. Both panels use hyperbolic arcsine intensity scaling. In green, we show contours of our computed lensing amplification model (in magnitudes), described in Section~\ref{sec:lensinginterp}.}
\label{fig:postagestamp}
\end{centering}
\end{figure*}

On 2016 Feb~29~UTC, we searched images of the WISE-selected \citep{wright10} massive galaxy cluster MOO~J1014+0038 ($z = 1.23$, \citealt{decker16}; $M_{200} = (5.6 \pm 0.6) \times 10^{14} M_{\odot}$, \citealt{brodwin15}), from the Massive and Distant Clusters of WISE Survey (MaDCoWS; \citealt{gettings12}, \citealt{stanford14}, \citealt{gonzalez15}). We detected a red (WFC3 $F140W = 25.1$, $F105W-F140W = 1.4$~AB~mag; upper limit only in $F814W$) supernova at $\alpha=153\fdg 52655$, $\delta= +0\fdg 64041$ (J2000, aligned to USNO-B1).  This SN was internally designated as SN~SCP16C03,\footnote{Nicknamed ``Joseph.''} as it was the third SN found in this cluster field (alphabetically labeled cluster ``C'') in 2016. As shown in Figure~\ref{fig:postagestamp}, the SN lies on a red galaxy having a color of $F105W-F140W = 1.5$~AB mag and is located 0\farcs 7 from the core of this galaxy ($\alpha=153\fdg 52643$, $\delta= +0\fdg 64025$). We conclude that this galaxy must be the host galaxy; there are no other galaxies nearby, the light curve is incompatible with those possible from an intra-cluster SN within MOO~J1014+0038 (as discussed in Section~\ref{sec:class}), and the probability of a chance projection is negligible ($\sim 0.2$\%, as discussed in Appendix \ref{sec:proj}).

\section{Spectroscopic and Photometric Follow-up}

\begin{figure}[h]
\begin{centering}
\includegraphics[width= 0.25 \textwidth]{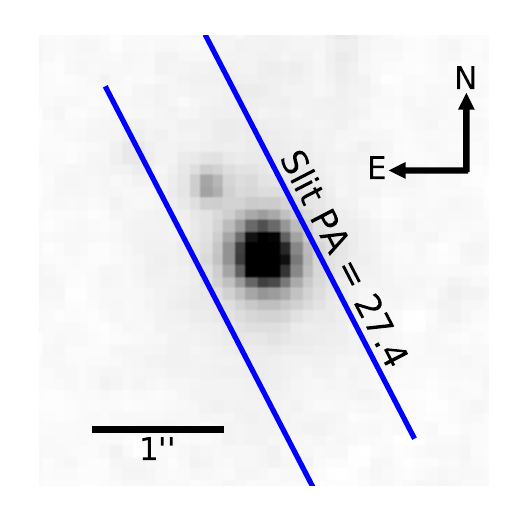}
\caption{Illustration of the X-shooter slit orientation and location on the SN and host galaxy.}
\label{fig:slit}
\end{centering}
\end{figure}

\begin{figure}[tbp]
\begin{centering}
\includegraphics[width=1.0\textwidth ]{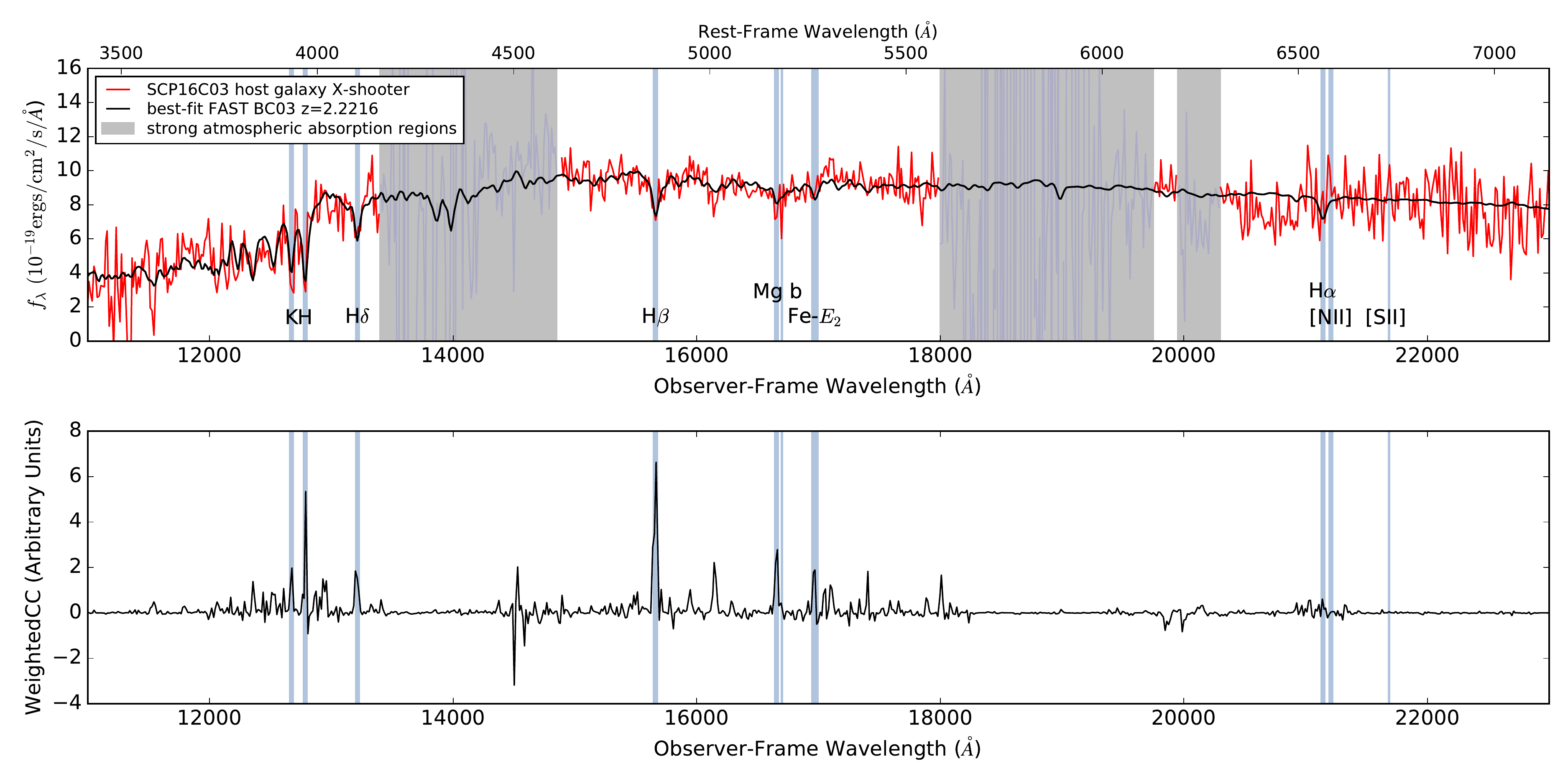}
\caption{\textit{Top:} Flux-calibrated, telluric corrected, de-lensed (by a factor of 2.8; see Section \ref{sec:lcfit}) and binned (14\AA\xspace observer-frame) spectrum of the host galaxy (red line). Gray shaded regions indicate strong atmospheric absorption. The best-fit \texttt{FAST} model (fit to both the photometry and spectrum) is shown in black, convolved by the 305 km/s velocity dispersion estimated by \texttt{ppxf}. Prominent features identified in both spectra are highlighted with light blue shading, and labeled; the H$\alpha$ line is labeled for reference, but no prominent contribution to the cross-correlation is detected from it. \textit{Bottom:} The contribution to the weighted cross-correlation of each wavelength element of the spectrum. Regions of heavy atmospheric absorption are properly de-weighted, and the prominent features easily identified by eye provide a confident redshift determination of $\redshift \pm 0.0002$.}
\label{fig:spectrum}
\end{centering}
\end{figure}

We activated ToO spectroscopic observations (PI: Hook) using X-shooter, the multi-wavelength medium resolution spectrograph on the Very Large Telescope \citep{vernet11}. Five ``Observation Blocks'' were taken between 2016 Mar~4-5 (UT), yielding a total integration time of 3.8, 4.4, and 5.0~hours in the X-shooter UVB, VIS, and NIR arms, respectively. Although the slit orientation captured both the host galaxy and the SN (shown in Figure~\ref{fig:slit}), the galaxy was $\sim 4$ magnitudes brighter so only a host-galaxy spectrum could be extracted. The data were reduced to wavelength- and flux-calibrated, sky-subtracted, 2D spectra using the Reflex software \citep{2013A&A...559A..96F}. Optimal 1D-extractions were determined using a combination of custom and \texttt{IRAF} \citep{1993ASPC...52..173T} routines.\footnote{IRAF is distributed by the National Optical Astronomy Observatory, which is operated by the Association of Universities for Research in Astronomy, Inc., under cooperative agreement with the National Science Foundation.} The extractions used a Gaussian profile in the cross-dispersion direction, the width of which was determined by fitting the trace in each (binned) 2d spectrum individually. These extractions were combined using the weighted mean to create a final 1D spectrum. We corrected for telluric absorption using a model atmosphere computed by the Line-By-Line Radiative Transfer Model \citep{clough92, clough05} retrieved through Telfit \citep{gullikson14}. Our observations were obtained at a low mean airmass of $\sim 1.15$ with a NIR slit width of $0\farcs 9$ (matching the mean seeing). The X-shooter NIR arm is aligned with the optical at 13,100 \AA\footnote{X-shooter user manual, Section 2.2.1.7 \url{http://www.eso.org/sci/facilities/paranal/instruments/xshooter/doc/VLT-MAN-ESO-14650-4942_v87.pdf}}; the atmospheric differential refraction over the NIR wavelength range for this airmass range is $< 0\farcs 1$. For a Gaussian PSF (with FWHM $0\farcs 9$) shifted by $0\farcs 1$, the differential slit loss would be $2\%$. Thus, atmospheric differential refraction has a negligible impact on the calibration of our NIR spectrum. The host-galaxy spectrum is shown in Figure~\ref{fig:spectrum}. Balmer H$\delta$ and H$\beta$ absorption lines are clearly detected, as are the Ca~H\&K absorption lines. Mg~b is likely detected as well. These lines provide the main contribution to the weighted cross-correlation, as shown in the lower panel of Figure~\ref{fig:spectrum}, yielding a redshift of $\redshift \pm 0.0002$ for the host galaxy. No emission lines were detected (see Section~\ref{sec:host:sfr} for further details). 

This SN was \supernovaBGCoffset away from the center of a dense concentration of cluster galaxies. Our initial flux amplification estimate was \initmagnification, calculated by assuming that the optical and halo center were the same, and employing a Navarro-Frenk-White dark matter profile \citep{nfw97}. Based on the likelihood that this would turn out to be a highly magnified, high-redshift SN~Ia, we applied for Director's Discretionary time on HST and were granted a three-orbit disruptive ToO to ensure a good measurement of the SN SED. We triggered $F105W$, $F125W$, and $F160W$ imaging from our ``See Change'' allocation. We did not request \HST grism spectroscopy, as the roll angle range available would not have allowed for a clean separation of the SN and its host. In addition to our triggered followup, we obtained additional WFC3 $F814W$, $F105W$, and $F140W$ imaging at two more previously scheduled orbits (the cadenced search visits for discovering SNe). We also were granted Director's Discretionary time on VLT for imaging in the $K_s$ band (PI: Nordin) with HAWK-I \citep{kissler08} in order to extend the wavelength range redder than is possible with \HST.

\clearpage
\section{SN Photometry}\label{sec:photclass}

With a pixel size of $\sim 0\farcs 128$, the WFC3 IR channel undersamples the PSF, making it difficult to remove underlying galaxy light with image resampling and subtraction. In addition, this SN poses a (unusual, but not unknown) challenge, as we have no reference (SN-free) images in $F125W$ and $F160W$, necessitating interpolation of the galaxy light over the SN position. To meet these needs, we have developed a ``forward modeling'' code described in \citet{suzuki12} and \citet{rubin13} and updated for WFC3 IR in \citet{nordin14}. This code fits each pixel as observed (flat-fielded ``flt'' images for WFC3 IR, CTE-corrected, flat-fielded ``flc'' images for WFC3 UVIS), i.e., without resampling due to spatial alignment or correction for distortion. It does this by creating an analytic model of the scene, convolving that model with the PSF, and fitting it to the pixel values. In all cases, the effect on the flatfield of pixel area variations over the FoV are taken into account in order to properly account for spatial distortion.

For the host-galaxy, our analytic model is a linear sum of an azimuthally symmetric spline (a 1D, radially varying component that is allowed to have ellipticity) and a 2D grid to capture azimuthal asymmetries.\footnote{The spline nodes were spaced less than 1 pixel ($0\farcs 128$) apart near the core for the 1D radially varying spline, and 4 pixels apart for the 2D spline. Reasonable variations around these values returned virtually the same photometry.} To these splines, we include a model for a point source representing the SN, with a different fitted amplitude in each epoch. As there are reference images in the cadenced IR filters ($F105W$ and $F140W$), these can be used as a cross-check of our ability to do photometry in $F125W$ and $F160W$, for which references are lacking. We obtain nearly identical photometry in the $F105W$ and $F140W$ bands with and without modeling the reference images, although the uncertainties are smaller when including references. Testing with simulated SNe indicates no photometric biases at the 0.01 magnitude level, and accurate uncertainties,\footnote{We use constant PSFs that do not follow \HST focus changes. We allow for these changes by adding 0.02 magnitudes (our measured dispersion on bright stars) in quadrature to the fit uncertainties.} even without using reference images. 

There is only weak evidence for SN light in the $F814W$ filter, as expected for such a high redshift 
SN~Ia. The low S/N necessitates a different treatment than the IR, where the offset between the SN and host galaxy was treated as a fit parameter. To determine the location of the SN in the $F814W$ images, we start by aligning the UVIS and IR exposures using \texttt{TweakReg},\footnote{All of the software tools referenced here are part of \texttt{DrizzlePac}: \url{http://drizzlepac.stsci.edu}.} and then drizzling to a common frame using \texttt{AstroDrizzle}. (These resampled images are not used for the SN photometry.) Then we perform PSF photometry on the UVIS images using the centroid transferred (using \texttt{pixtopix}) from that measured in the IR images, where the SN is detected. We fit for a spatially constant background underneath the SN, incorporating both sky and underlying galaxy. Even the core of the galaxy is only detected at low S/N in each $F814W$ exposure, so the second derivative\footnote{As the fit patch is centered on the SN, any slope in the underlying galaxy light will cancel out. Thus, only the second derivative of the light can affect the photometry, and the second derivative at the SN location is much smaller than at the galaxy core.} underneath the SN is small.

Bad weather allowed only one of the seven Hawk-I $K_s$ observing blocks to be executed; 40 minutes of effective exposure time was obtained in excellent seeing (0\farcs 27) enabling a measurement of the SN flux. We use the same host-galaxy model as for the WFC3 IR data, but with the offset of the SN from the host fixed, and employing field stars to derive the PSF. We determine the flux calibration (in AB mags) from the standard star FS19 and from the 2MASS \citep{skrutskie06} magnitudes of the field stars.   Although the brightness uncertainty of 60\% is too large to improve the distance measurements, the additional rest-frame $R$-band coverage helped with photometric classification.

We present the resulting SN photometry in Table~\ref{tab:photometry}. The uncertainty in the host-galaxy subtraction in each band leads to correlations between the WFC3 IR photometry measurements (in the data for each filter), 
so we also provide the inverse covariance matrix in Table~\ref{tab:weight}. We take the WFC3 IR zeropoints from \citet{nordin14}, 
converted to AB magnitudes, which take into account the WFC3 IR count-rate nonlinearity. For the UVIS F814W filter, we use the STScI-provided UVIS zeropoint, after correcting the encircled energy to an infinite aperture.

\begin{deluxetable}{ccccccc}
\tablecolumns{4} 
 \tablecaption{Photometry for SN SCP16C03.
 \label{tab:photometry}}
 
 \tablehead{
 \colhead{MJD} & \colhead{UT} &  \colhead{Observation} & \colhead{Flux (e$^-$/s)} & \colhead{Uncertainty} & \colhead{Chip} & \colhead {AB Zeropoint}}
 
 \startdata 
 \cutinhead{\HST $F814W$}
57414.741 & 01-27-2016 & Cadenced & 0.162 & 0.079 & 2 & 25.146 \\
57446.809 & 02-28-2016 & Cadenced & 0.079 & 0.070 & 1 & 25.146 \\ 
57479.413 & 04-01-2016 & Cadenced & 0.056 & 0.071 & 2 & 25.146 \\
57513.068 & 05-05-2016 & Cadenced & 0.000 & 0.066 & 1 & 25.146 \\
 \cutinhead{VLT $K_s$}
57479.051 & 04-01-2016 & Triggered & 173 &  102 & 3 & 30.26 \\
\cutinhead{\HST $F105W$}
57414.751 & 01-27-2016 & Cadenced & 0.095 & 0.145 & 1 & 26.235 \\
57446.824 & 02-28-2016 & Cadenced & 0.779 & 0.197 & 1 & 26.235 \\
57465.544 & 03-18-2016 & Triggered & 1.543 & 0.094 & 1 & 26.235 \\
57479.426 & 04-01-2016 & Cadenced & 1.542 & 0.161 & 1 & 26.235 \\
57513.081 & 05-05-2016 & Cadenced & 0.483 & 0.156 & 1 & 26.235 \\
\cutinhead{\HST $F125W$}
57462.864 & 03-15-2016 & Triggered & 3.205 & 0.172 & 1 & 26.210 \\
57465.506 & 03-18-2016 & Triggered & 3.351 & 0.173 & 1 & 26.210 \\
\cutinhead{\HST $F140W$}
57414.756 & 01-27-2016 & Cadenced & 0.363 & 0.161 & 1 & 26.437 \\
57446.820 & 02-28-2016 & Cadenced & 3.294 & 0.245 & 1 & 26.437 \\
57479.413 & 04-01-2016 & Cadenced & 4.835 & 0.213 & 1 & 26.437 \\
57513.068 & 05-05-2016 & Cadenced & 2.396 & 0.220 & 1 & 26.437 \\
\cutinhead{\HST $F160W$}
57462.909 & 03-15-2016 & Triggered & 2.673 & 0.176 & 1 & 25.921 \\
57465.572 & 03-18-2016 & Triggered & 2.691 & 0.182 & 1 & 25.921 \\
\enddata
 \tablecomments{Photometry measured for each band on each date. Table~\ref{tab:weight} has the weight matrices for the WFC3 IR measurements.}
\end{deluxetable}

\begin{deluxetable}{lrrrrr}
\tablecolumns{6} 
 \tablecaption{Inverse covariance matrices for the WFC3 IR photometry.
 \label{tab:weight}}
 
 \tablehead{
 \colhead{MJD} & \multicolumn{5}{c}{Inverse Covariance Matrix}}
 \startdata 
 \cutinhead{$F105W$}
57414.751  &  $48.2472$  &  $-1.2956$  &  $-5.6335$  &  $-1.7874$  &  -1.8504  \\
57446.824  &  $-1.2956$  &  $26.0744$  &  $-3.3404$  &  $-1.0842$  &  -1.1516  \\
57465.544  &  $-5.6335$  &  $-3.3404$  &  $115.3885$  & $-4.6158$  &  -4.7575  \\
57479.426  &  $-1.7874$  &  $-1.0842$  &  $-4.6158$  &  $39.0092$  &  -1.5386  \\
57513.081  &  $-1.8504$  &  $-1.1516$  &  $-4.7575$  &  $-1.5386$  &  41.7656  \\
 \cutinhead{$F125W$}
57462.864  &  $35.4715$  &  $-7.4421$  &  \nodata &  \nodata &  \nodata \\
57465.506  &  $-7.4421$  &  $35.0010$  &  \nodata &  \nodata &  \nodata \\
 \cutinhead{$F140W$}
57414.756  &  $39.2367$  &  $-1.5114$  &  $-1.8203$  &  $-1.8066$  &  \nodata   \\
57446.820  &  $-1.5114$  &  $16.7848$  &  $-0.7807$  &  $-0.9239$  &  \nodata  \\
57479.413  &  $-1.8203$  &  $-0.7807$  &  $22.1654$  &  $-1.0547$  &  \nodata  \\
57513.068  &  $-1.8066$  &  $-0.9239$  &  $-1.0547$  &  $20.8010$  &  \nodata  \\
 \cutinhead{$F160W$}
57462.909  &  $34.5919$  &  $-8.5948$  &  \nodata &  \nodata &  \nodata  \\
57465.572  &  $-8.5948$  &  $32.1690$  &  \nodata &  \nodata &  \nodata \\
\enddata

 \tablecomments{Inverse covariance matrices for the WFC3 IR photometric measurements. These data are correlated (within each band) due to uncertainty in the underlying host-galaxy light.}
\end{deluxetable}
\clearpage

\section{Host Galaxy Properties}\label{sec:host}

In this section, we present our measurements of the host-galaxy properties from spectroscopy and imaging, and summarize the evidence that the galaxy is old for its redshift and therefore lacking significant star formation in Section~\ref{sec:host:summary}.

\subsection{Host-galaxy Photometry}\label{sec:host:phot}

We perform photometry of the host galaxy on aligned, stacked (SN-free) images using \texttt{SEP} \footnote{SEP is a Python implementation of SExtractor Bertin \& Arnouts \url{https://sep.readthedocs.io}} To minimize contamination by light from the nearby galaxies, we use an elliptical aperture set to 1.2 times the Kron radius \citep{kron80}.\footnote{The major and minor radii are 1\farcs19 and 0\farcs83, respectively.} In addition to the \HST WFC3 photometry, we use archival Spitzer IRAC CH1 and CH2 data \citep{gonzalezInPrep}, and we obtained bluer $g'$, $r'$ and $i'$ photometry of the host with the OSIRIS instrument at the Gran Telescopio Canarias (GTC). We determine the zeropoints of this photometry using an SDSS star in the GTC field of view (both the star and galaxy are measured with the same elliptical aperture). We present the host-galaxy photometry in Table~\ref{tab:hostphotometry}. 

\begin{deluxetable}{ccrr}
\tablecolumns{4} 
 \tablecaption{Host-galaxy photometry.
 \label{tab:hostphotometry}}
 \tablehead{
 \colhead{Instrument} & \colhead{Filter} & \colhead{Flux ($\mu$Jy)} & \colhead{Flux Uncertainty ($\mu$Jy)}}
 \startdata 
OSIRIS & $g'$ & 0.19 & 0.03 \\
OSIRIS & $r'$ & 0.37 & 0.06 \\
OSIRIS & $i'$ & 0.55 & 0.10 \\
WFC3 & $F814W$ & 1.06 & 0.20 \\
WFC3 & $F105W$ & 3.66 & 0.08 \\
WFC3 & $F125W$ & 8.72 & 0.32 \\
WFC3 & $F140W$ & 15.05 & 0.08 \\
WFC3 & $F160W$ & 20.79 & 0.44 \\
Hawk-I & $K_s$ & 32.09 & 1.12 \\
IRAC & CH1 & 57.1 & 1.7 \\
IRAC & CH2 & 63.8 & 1.7 \\
\enddata
 \tablecomments{Measured host-galaxy fluxes in the \HST and GTC imaging. We scale to $\mu$Jy (AB zeropoint of 23.9).}
\end{deluxetable}

\subsection{Host Age, Star-formation Rate, and Mass from SED Fitting}\label{sec:host:sed}

While SED fitting to broadband photometry has a well-known degeneracy between age, dust, and metallicity \citep[e.g.][]{bell01}, the X-shooter spectrum has many absorption features that can help break that degeneracy. To simultaneously fit the spectrum and the photometry, we use \texttt{FAST} \citep{kriek09}.

We calibrate the X-shooter spectrum to the de-lensed HST photometry in bands $F125W$, $F140W$, and $F160W$ by integrating the de-lensed X-shooter spectrum over these filters; we find a multiplicative factor of 1.5 in flux is required to match the \HST photometry. We then bin the X-shooter spectrum to 5~\AA. We use the high-resolution stellar population synthesis models included with \texttt{FAST}, performing a fit using both \citealt{bruzual03}  (BC03) and \citealt{con:fsps1} (FSPS) models, and averaging the two results (they agree extremely well). We use the Chabrier IMF \citep{chabrier03}, with a delayed-exponential star-formation history and the default `kc' parameterization of the dust attenuation curve \citep{kriek13}. We first fit to a coarse grid, then progressively fit to a finer grid near the best-fit region of parameter space to better estimate uncertainties. Note that \texttt{FAST} does not model emission lines. \texttt{FAST} returns a $\chi^2$ of 1385 for 2182 degrees of freedom for the BC03 models, and $\chi^2$ of 1414 for 2179 degrees of freedom using the FSPS models. The final uncertainties are estimated using the \texttt{FAST} Monte Carlo simulation option, sampling 1000 times. The best-fit model and derived parameters are presented in Table~\ref{tab:hostprop2} and the model spectrum is shown overlaid on the X-shooter spectrum in Figure~\ref{fig:spectrum}. The best-fit composite-stellar-population template has an age of \fastage~Gyr; as the age of the universe at $z=\shortredshift$ is only 2.96~Gyr, the implied formation redshift is $z_{\mathrm{form}} \sim 4.5$. The galaxy is very massive, with a stellar mass of \fastmass~M$_{\odot}$, and it has a low SFR of \FASTSFR~M$_\odot$/yr (averaged over the prior 30 Myr) given its mass. The \texttt{FAST} star-formation history model indicates that 99.94\% of all stars expected to ever form have already formed in this galaxy. \texttt{FAST} finds little dust, with a mean $A_V = $~\fastAv. We assume a two-component dust model of \citet{charlot00} where the dust optical depth for older stars is 0.3 times the optical depth for young (i.e., $<10$ Myr) stars. This implies $A_V \sim \fastAvthree$ mag for any young stellar component.  The age measurement is sufficiently precise to contribute a 
cosmochronometry measurement and thereby help constrain cosmological parameters \citep[e.g.,][]{jimenez02, stern10, moresco12}.

\subsection{Emission-line measurement of the Star-Formation Rate}\label{sec:host:sfr}

\begin{figure*} 
\begin{centering}
\includegraphics[width= 0.45\textwidth]{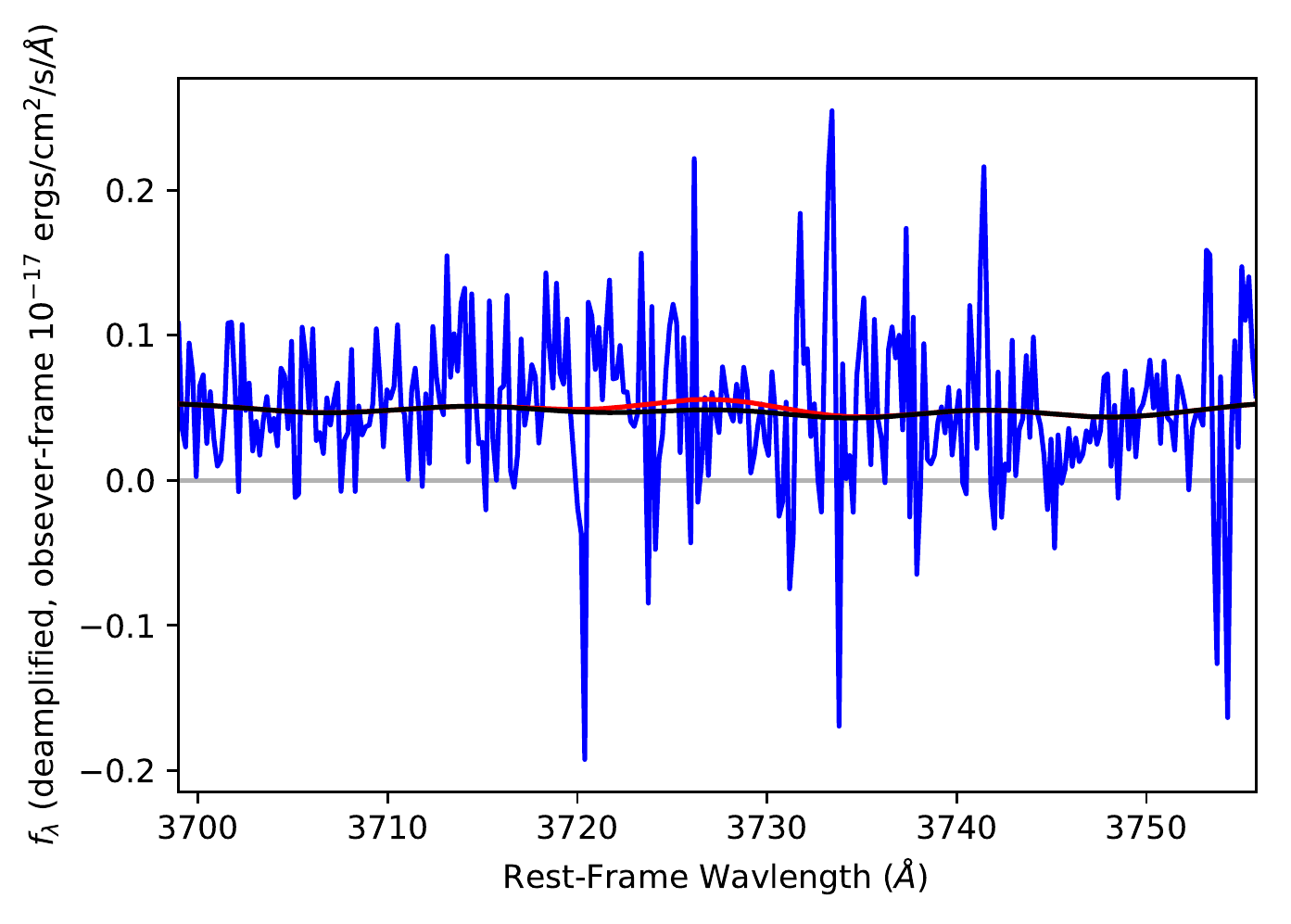}
\includegraphics[width= 0.45\textwidth]{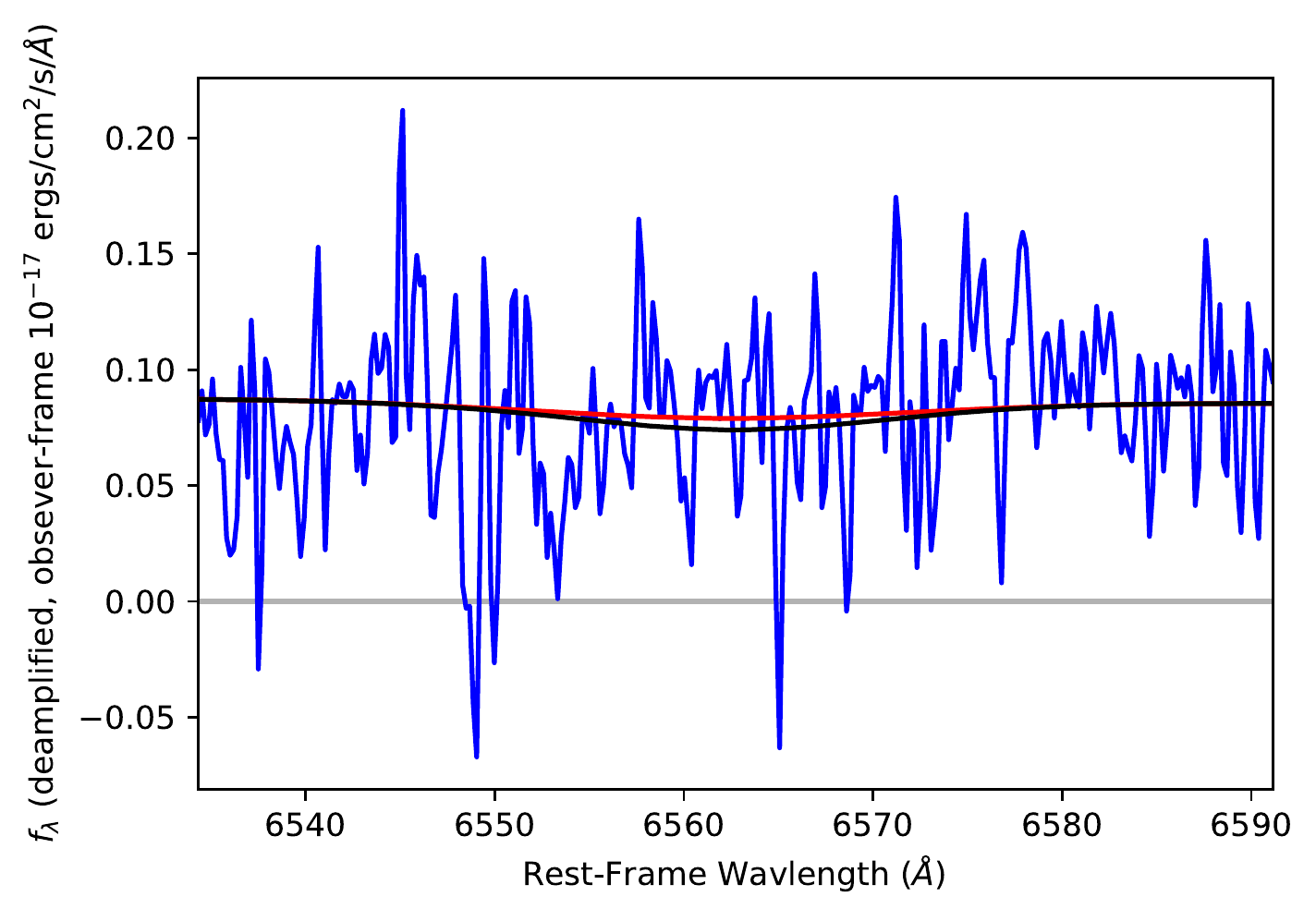}

\caption{In blue, we show cutouts of the spectral regions for [OII] (left panel) and H$\alpha$ (right panel). No emission lines are detected in either region. In black, we show the best-fit \texttt{FAST} template, convolved to the \texttt{ppxf} velocity dispersion of \ppxfvdispnoerror. In red, we show the best-fit [OII] (left panel) and H$\alpha$ (right panel) contribution.}
\label{fig:EW}
\end{centering}
\end{figure*}

We measure equivalent widths from the spectral extractions centered on the core of the galaxy (shown in Figure~\ref{fig:EW}). The region around [OII] and H$\alpha$ are modeled with the sum of the best-FIT \texttt{FAST} SED model (convolved to the \texttt{ppxf} velocity dispersion of \ppxfvdispnoerror) and Gaussian emission. We fix the emission component to have a velocity dispersion of \ppxfvdispnoerror. For [OII], we use two Gaussians, with the relative amplitude fixed to 0.74 for the $3726$\AA\xspace component and 0.26 for the $3729$\AA\xspace one.

We find a [OII]~$\lambda 3727$\,\ang rest-frame equivalent width of $\textrm{EW([OII])} =\oiimean$~\AA, as shown in Figure~\ref{fig:EW}. We see no evidence of [OIII]~$\lambda 5007$\,\ang emission; a nearby strong night sky line residual prevents us from measuring this spectral region. After correcting for the stellar Balmer absorption predicted by the \texttt{FAST} SED fit, we measure H$\alpha$ having EW(H$\alpha$)~$=\halphamean$~\ang. These weak emission line equivalent widths point to a low SFR per unit stellar mass.\footnote{As a cross-check, we also perform an extraction centered at the SN location, using the same extraction profile weighting as derived for the galaxy extraction. This location is not fully independent, as it is $0\farcs7$ offset from the core, and the seeing was about $0\farcs9$. For this more local measurement, we marginalize the width of the emission component from 30 to 300 km/s (with a flat prior in this range) and the velocity offset (with a Gaussian prior of 300 km/s), as it may be localized in the galaxy. We again find no evidence of emission.}

Using the X-Shooter calibration discussed in Section~\ref{sec:host:sed}, and correcting for extinction, we convert our H$\alpha$ equivalent width to a line luminosity of $L(\mathrm{H}\alpha) = \halphalum$~ergs/s.\footnote{The cosmology does not matter much here, but we use a flat $\Lambda$CDM cosmology with $\Omega_m = 0.3089$ \citep{planck15}. Note that we also must account for possible additional extinction of the H$\alpha$ region compared to the older stars in the rest of the galaxy, see Section~\ref{sec:host:sed}. We thus use $A_V$ = \fastAvthree, rather than the \texttt{FAST} result of \fastAv.}
We then convert to a star formation rate using $\dot{\mathrm{M}}=5.45\times 10^{-42}\, L(\textrm{H}\alpha)$~(M$_\odot$/yr)/(erg/s) from \citet{calzetti10}, resulting in $\dot{M} = \HalphaSFR$~M$_{\odot}$/yr. The H$\alpha$ and SED-fit star-formation rates agree well, within $\halphaSEDagree\,\sigma$.
At low SFR, OII emission can be badly contaminated 
by emission from post-AGB stars \citep[e.g.,][]{belfiore16}. If there were no such contamination, the derived SFR would be \OIISFR~(M$_\odot$/yr)/(erg/s), using the conversion $\dot{M}=2.65\times 10^{-41}\, L(\textrm{OII})$ \citep{meyers12} and our \texttt{FAST} SED extinction-corrected [OII]~$\lambda 3727$\,\ang luminosity of L([OII]) \oiilum M$_{\odot}$/yr)/(erg/s). This SFR estimate is consistent with the others (although with much larger uncertainties). However, we do not use it to estimate the SFR for this host. Likewise, while [OIII]~$\lambda 5007$\,\ang is usually present when there is star formation, it is a poor SFR indicator due to its strong dependence on metallicity and the hardness of the stellar UV spectrum.

\begin{figure*} 
\begin{centering}
\includegraphics[width= 0.90\textwidth]{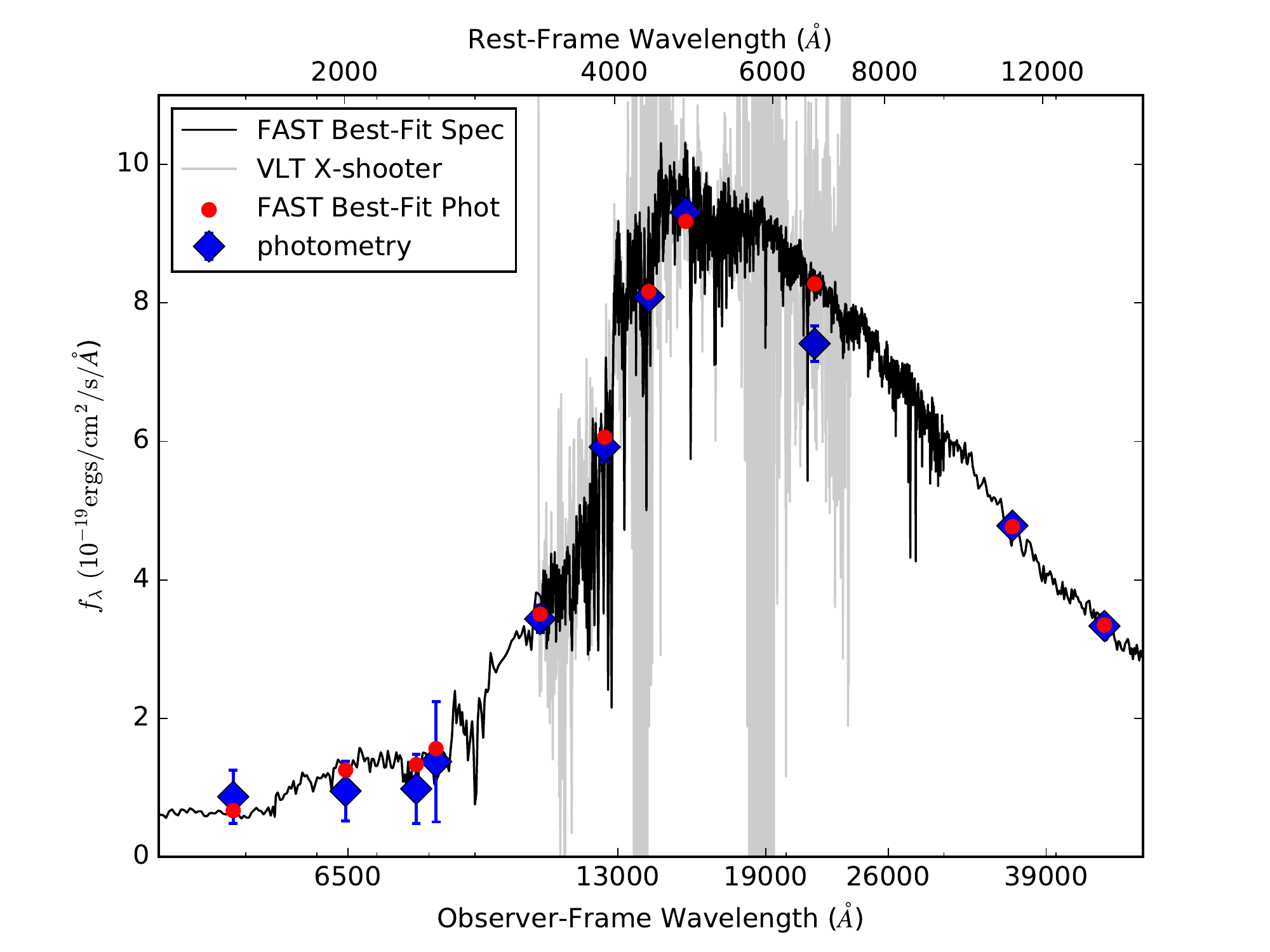}
\caption{Host galaxy photometry and spectrum with the best-fit \texttt{FAST} model. The HST photometry was used for flux calibration of the X-shooter spectrum. The best-fit \texttt{FAST} model for both BC03 and FSPS templates is a massive galaxy with low SFR \FASTSFR M$_\odot$/yr. The age, SFR timescale parameter $\tau$, and rest-frame $U-V$ and $V-J$ colors (well-constrained by the GTC, HST, and Spitzer photometry) all imply a quiescent galaxy; with an age that is 10 times the e-folding timescale of the best fit stellar population model, 99.94\% of stars expected to ever form in this galaxy have already formed.}
\label{fig:SED}
\end{centering}
\end{figure*}

\begin{deluxetable}{lcl}
\tablecaption{SCP16C03 Host-galaxy Properties\label{tab:hostprop2}}
\tablehead{\colhead{Quantity} & \colhead{Value} & \colhead{Notes}}
\startdata
EW(OII~3727) (\AA)                           & \oiimean        & \\
EW(H$\alpha$) (\AA)                          & \halphamean     & corrected for stellar absorption \\
Emission line SFR (M$_\odot$/yr)             & \HalphaSFR      & corrected for \texttt{FAST} $A_V$\\
Galaxy SED SFR (M$_\odot$/yr)               & \FASTSFR      & from \texttt{FAST} SED fit -- integrated over past 30~Myr \\
$\log_{10}$(M/M$_\odot$)                     & \fastmass       & from \texttt{FAST} SED fit \\
$\log_{10}$(Specific SFR (1/yr))             & \SSFR       & derived from best-fit \texttt{FAST} SED \\
age (Gyr)                                    & \fastage        & age of universe $2.96$ Gyr at $z=2.2216$ \\
$\tau$ (Gyr)                                 & \fasttau        & SFR $\propto$ $t$e$^{-\mathrm{t}/\tau}$ \\
$A_V$                                        & \fastAv         & using `kc' dust parametrization in \texttt{FAST} \\
$(U-V)_{\mathrm{AB}}$                        & 1.74            & rest-frame $U-V$ derived from best-fit \texttt{FAST} SED \\
$(V-J)_{\mathrm{AB}}$                        & 0.9             & rest-frame $V-J$ derived from best-fit \texttt{FAST} SED \\
$\sigma$ (km/s)                              & \ppxfvdisp      & velocity dispersion measured with \texttt{ppxf}\tablenotemark{a} \\
$n$                                          & \nsersic        & \sersic index from forward model \\
r$_{\mathrm{eff}}$ (arcsec)                  & \halflight      & half-light radius from raw pixels; not de-magnified \\
r$_{\mathrm{eff}}$ (kpc, de-mag)             & \halflightdemag & half-light radius in kpc, de-magnified by 2.8$\times$ \\
log$_{10}\Sigma$ (M$_\odot /\mathrm{kpc}^2$) & \centraldensity & derived compactness \\
\enddata
\tablenotetext{a}{\citet{ppxf04,ppxf17}}
\end{deluxetable}

\subsection{Profile Fitting}\label{sec:host:sersic}

\begin{figure*}
\begin{centering}
\includegraphics[width= 0.8\textwidth]{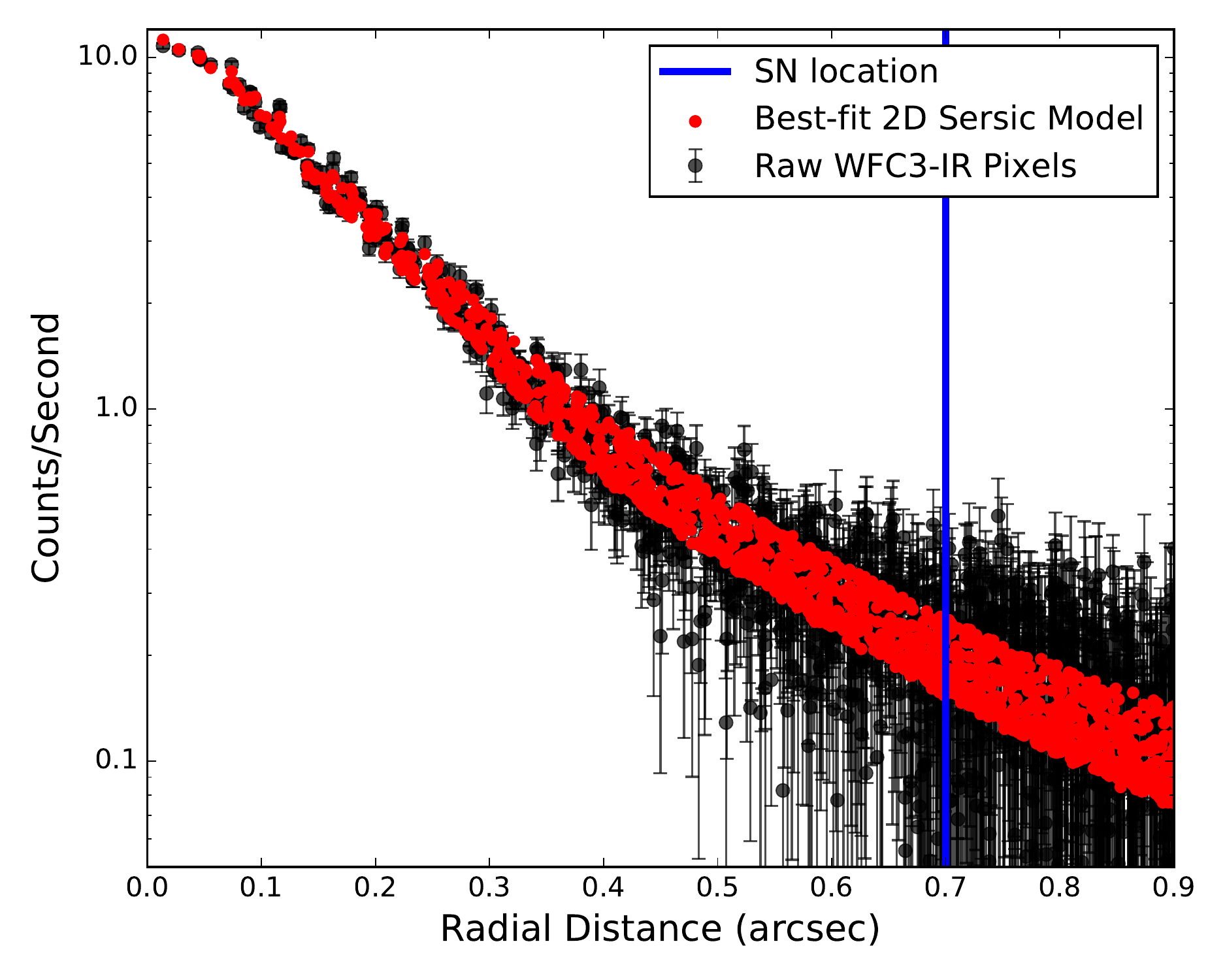}
\caption{The radial brightness profile of the host galaxy (black points), with the best-fit 2D \sersic model (red points). The scatter in the best-fit \sersic model is a combination of the elliptical galaxy plotted in radial coordinates, and the asymmetric WFC3 IR PSF. The galaxy is well-fit by the 2D \sersic profile, with a best-fit \sersic index of $2.8\pm 0.1$
and a half-light radius of \halflight. The 
galactocentric radius at which the SN occurred is indicated by a vertical blue line at $0 \farcs 7$.}
\label{fig:sersic_radial_profile}
\end{centering}
\end{figure*}

We also measure the surface brightness profile of the host galaxy. We fit the $F140W$ imaging, as it is close to rest-frame $B$-band. To determine the central profile given the undersampled imaging, we again use a forward-model approach. We model the observed pixels (without resampling) of nine dithers (from SN-free epochs) with a \sersic profile \citep{sersic63} convolved with the PSF (the same PSF as for the SN photometry). We marginalize over the amplitude, half-light radius, ellipticity, and orientation\footnote{Marginalizing over the ellipticity also has the effect of making the fit less sensitive to lensing shear.}. The galaxy is fit well by this simple model, with no evidence of an on-core point source (or variability) which might imply a central AGN (see Figure~\ref{fig:sersic_radial_profile}). We find a \sersic index of $2.8 \pm 0.1$, and a de-magnified half-light radius of \halflightdemag~kpc. In combination
with the total stellar mass, this implies central density of log$_{10}\Sigma$ (M$_\odot /\mathrm{kpc}^2$) = \centraldensity, as summarized in Table~\ref{tab:hostprop2}. 

\subsection{Host Galaxy Summary} \label{sec:host:summary}

All host parameters point to the host galaxy being a massive quiescent galaxy. The velocity dispersion from the spectrum and the stellar mass from SED fitting provide joint confirmation of a high mass, while the SED fit to the spectrum and photometry provides a strong constraint on low SFR, reinforced by lack of evidence for emission lines in the spectrum. The derived specific SFR of log$_{10}(\mathrm{sSFR})=\SSFR$ qualifies this galaxy as being in a stage of quiescent star-formation, as defined by the mass-doubling criterion of \citet{feulner05} or the log$_{10}(\mathrm{sSFR}) < -11$ criterion of \citet{dominguez11}. 
 The 10 star-formation $e$-folding times that have passed reinforces this conclusion. The structure of the host galaxy -- its moderately high \sersic index, small $r_{\mathrm{eff}}$, and high central density -- is also consistent with other massive quiescent galaxies found at similar redshifts \citep{muzzin12,vanderwel14,newman15,vandokkum15,barro17}.

At these high-redshifts it has become standard practice to define quiescence photometrically, based on the statistical separation of galaxies in rest-frame $U-V$ versus $V-J$ \citep[the $UVJ$ diagram;][]{williams09, brammer11} into a red, dead ``clump'' and a star-forming ``track'' derived using the UV$+$IR SFR. Spectroscopic follow-up has confirmed that the $UVJ$ diagram selects older, quiescent galaxies efficiently \citep{whitaker13}. In this space, dusty star-forming galaxies move along a track that is perpendicular to the separation between the quiescent clump and the star-forming track \citep{williams09,brammer11}. Movement from the star-forming track to the quiescent clump is well-described by passive stellar evolution \citep{williams09}. The unpopulated region between the quiescent clump and the star-forming track is known as the the ``green valley''; it is not yet established whether these galaxies represent a short-lived (and dusty) evolutionary path from the star-formation main sequence to the red sequence of galaxies, or are merely the overlapping tails of two distinct populations with measurement scatter \citep[see e.g.][for a review]{pandya17}. Regardless, the $U-V$ and $V-J$ colors of the galaxy place it firmly in the quiescent regime according to the $UVJ$ classification scheme (see Table~\ref{tab:hostprop2} and e.g. \citealt{brammer11}).

The lack of emission lines that trace star-formation, the low specific SFR and stellar extinction, the red color of the SED from rest-frame $U$ to $J$, and estimates of morphology from the imaging combine to strongly imply a quiescent galaxy, and not a dusty star-forming galaxy.\footnote{In Appendix \ref{sec:obscuredCC}, we evaluate the feasibility of a starburst component added to the best-fit \texttt{FAST} model, finding the scenario to be unlikely, and with only a small amount of obscured star-formation allowed.} Of course very sharp-edged dust structures, such that the affected stars contribute negligibly to the light and associated extinction we do detect, could hide more stellar mass and star formation. Such dust would also be required to preserve the smooth structure of this galaxy. In this particular case, such a contrived means of hiding significant stellar mass is especially unlikely since this galaxy is already on the steeply falling segment of the galaxy luminosity function. For instance, if half the stellar mass were hidden, the $2.0<z<2.5$ galaxy luminosity function \citep[e.g.][]{davidson17} predicts a $65\times$ lower incidence per unit volume for the resulting galaxy. We also cannot rule-out a situation in which the specific line of sight for this SN has much less dust, thereby escaping these constraints from host galaxy properties. But such a finely-tuned case seems unlikely, and is unnecessary to explain our observations, as the following section will make apparent.

\section{SN Photometric Classification}\label{sec:class}

\begin{figure*}[tbp]
\begin{centering}
\includegraphics[width= 0.9 \textwidth]{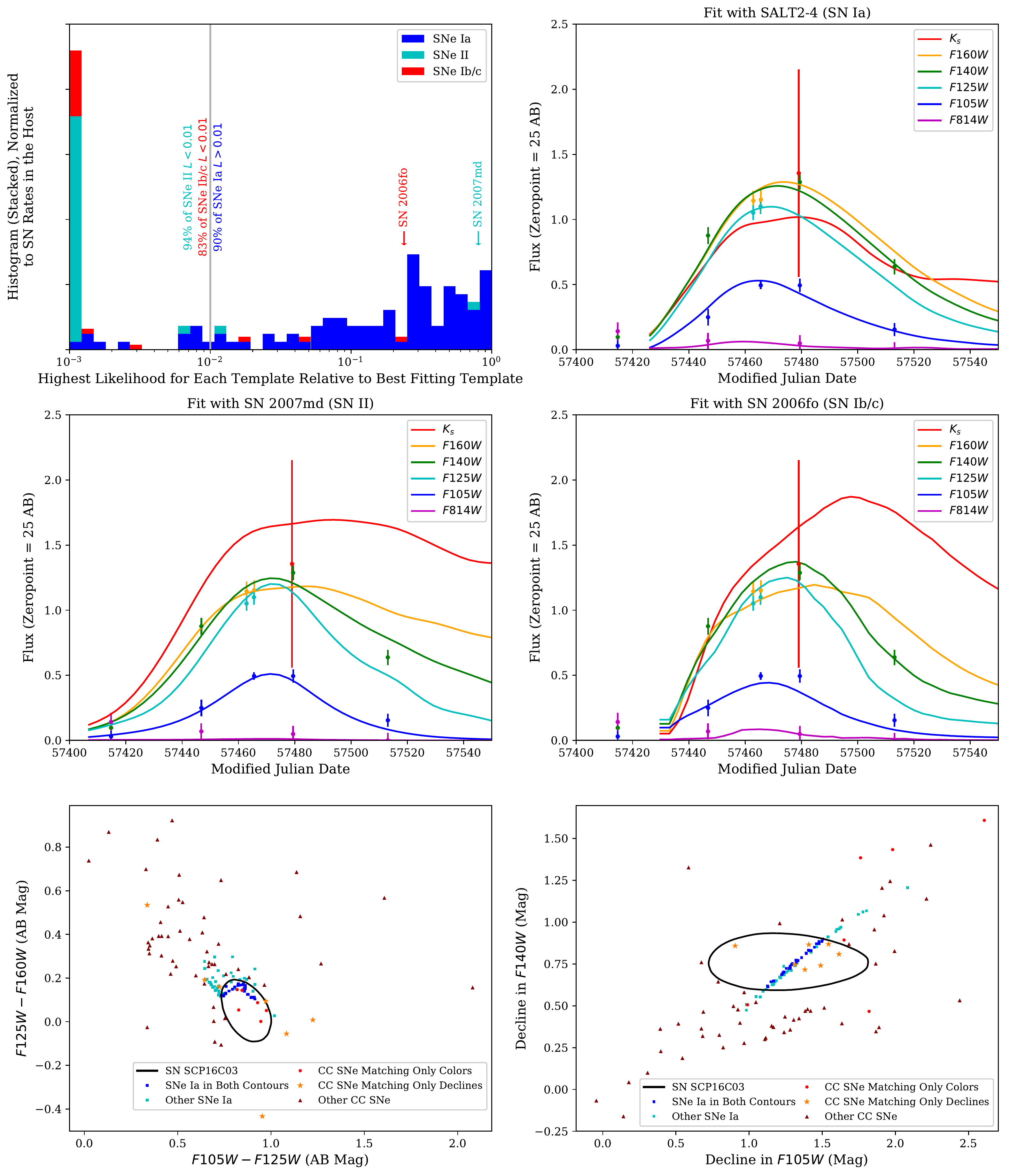}
\caption{Top left panel: histogram of relative likelihoods from the light curve fits to the templates of each type of SN. The likelihoods are scaled so that the template with the highest likelihood has a value of 1. The histogram is normalized such that the area for each type of SN matches the relative rates expected in this galaxy (Section~\ref{sec:class}). This histogram indicates that, although there are two CC templates with a reasonable fit to the light curves, there are many more that are poor fits and thus the probability that SN~SCP16C03 is a SN~Ia is high. Light-curve panels: the best-fit for the SN~Ia templates (top right), the SN~II templates (middle left), and the SN~Ib/c templates (middle right). In the bottom panels, we show how SN~SCP16C03 compares to the template library in both color (bottom left: the $F105W - F125W$ and $F125W - F160W$ colors at the epoch with all three) and decline rate (bottom right: the decline between the last two epochs in $F105W$ and $F140W$). Each point plotted is from the best-fit (i.e., best-fit amplitude, date of maximum, and extinction) of one of the SN templates (for display purposes, we show as many SNe~Ia realizations as  CC templates, although most SNe in the host galaxy are SNe~Ia). The black contours show the 68\% confidence intervals from our measurements of SN~SCP16C03. The blue squares represent SNe~Ia templates inside both contours; the cyan squares represent SNe~Ia outside of at least one contour. The consistency of SN~SCP16C03 with the SN~Ia distribution is evident. There are no CC SNe in both contours. The red dots represent CC SNe matching the colors, the orange stars represent CC SNe matching the observed declines, and the brown triangles represent CC SNe matching neither colors nor declines.}
\label{fig:lc}
\end{centering}
\end{figure*}

It was not possible to obtain a spectrum of the SN itself, so we use our photometry, host-galaxy redshift, and host-galaxy star-formation rate to ascertain the SN type. The major types we wish to distinguish are SNe~Ia and CC SNe. Within each of these are subtypes: SNe~II and SNe~Ibc among the CC SNe, and normal and abnormal subtypes among the SNe~Ia. In order to arrive at an estimate for the probability of each SN type to be the type for SN~SCP16C03, we must take into account several factors. The first is the average relative likelihood for each (sub)type, \Ltype, derived using the light curve fits. Then, we need an estimate of the {\it a priori} SN~Ia fraction, $f_{\mathrm{Ia}}$, for the specific host of SN~SCP16C03. An estimate of the {\it a priori} relative incidence of the normal vs abnormal subtypes for SNe~Ia and the SN~II and SN~Ibc subtypes for CC SNe that correspond to our light curve templates is also needed. The product of $f_{\mathrm{Ia}}$ with these subtype apportionments gives the {\it a priori} expectation, $f_{\type}$, for each type of template. These can then be combined to give the probability of each (sub)type based on our measurements:

\begin{equation}
\label{eq:type}
P_{\type} = \frac{f_{\type}\; \Ltype/\LtypeIa}{\sum f_{\type}\; \Ltype/\LtypeIa} 
\end{equation}

\subsection{Light-curve likelihoods}

Quantitative photometric typing is made more challenging by the limited signal-to-noise, wavelength coverage, and phase coverage of our light curves. Similarly to the photometric typing efforts performed for other high-redshift SNe \citep{rodney12, jones13, nordin14, patel14, rodney15}, as well as the Photometric Supernova Identifier \citep{sako11}, we fit the photometry with a range of templates to compare light curve shape and color. Non-parametric methods have shown comparable promise \citep[e.g.,][]{kessler10, lochner16}, but better temporal sampling than available here would be necessary to implement these. 

The core-collapse templates are taken from SNANA \citep{kessler09}, some of which are described in \citealt{gilliland99}, \citealt{nugent02}, \citealt{stern04}, \citealt{levan05}, and \citealt{sako11}, and are implemented in \texttt{sncosmo}\footnote{\url{http://sncosmo.readthedocs.io}}. When multiple versions of the same non-Ia template are available, we conservatively take that with the lowest $\chi^2$ (giving that non-Ia SN template the highest probability of matching). We do not expect the CC template set to be complete for several reasons. Typically, the absolute magnitudes of CC SNe are fainter than those of SNe~Ia \citep{richardson14}, so proportionately fewer are found. Furthermore, the CC SNe that are discovered are less likely to be followed up (because they are not as cosmologically useful).

For the SN~Ia templates, we construct a random sample with the latest version (SALT2-4) of the Spectral Adaptive Lightcurve Template model \citep{guy07,mosher14}  (SALT2 is described more in Section~\ref{sec:lcfit}), with Gaussian distributions of the light-curve shape parameters $x_1$ and $c$, centered at zero and having dispersions of 1 in $x_1$ and $0.1$ in $c$. Because SALT2 is a parameterized model, we use it to simulate a distribution of SNe~Ia. We also include the \citet{hsiao07} SN~Ia spectral template, the \citet{nugent02} templates for normal and underluminous SN~1991bg-like SN~Ia subtypes, and the \citet{stern04} template for the overluminous SN~1991T-like SN~Ia subtype. SALT2 does not mimic the SN~1991bg-like and SN~1991T-like subtypes well, so we separately keep track of those subtypes. The templates of normal SNe~Ia from \citet{hsiao07} and \citet{nugent02} are only used as a cross-check of our SALT2 results.

Next we fit the templates to the data. We leave the date of maximum and normalization (i.e., luminosity) fully unconstrained. As the unreddened colors of the templates are unknown, it is necessary to allow the relative extinction between SN~SCP16C03 and the SN used to determine the template colors to vary between positive and negative values. We achieve this using a broad, double-sided exponential prior having a scale length of 0.2 in $E(B-V)$. As $R_V$ cannot be constrained by the data, we fix it to the Galactic value of 3.1 \citep[e.g.,][]{schlafly16} for all SN types). It is also necessary to account for the fact that the parameter space is sampled with only a finite number of templates. As with previous work \citep{rodney09}, we add 0.15 magnitudes uncertainty in quadrature with each photometry point to address this.

The three best-fitting templates (with a spread in log-likelihood of just 0.2) are the SALT2 SN~Ia template (close to the mean of the $x_1$ and $c$ distributions), the Nugent template for SN~1991T-like SNe~Ia, and an SDSS SN~II (SN~2007md). Encouragingly, the best-fit SALT2 template has a $\chi^2$ of 8.7 for 8 DoF (including the SALT2 model uncertainties, but not the 0.15 magnitude uncertainty floor). Other templates with a best-fit likelihood within a factor of 20 of the SN~Ia template are the Nugent SN~Ia template, an SDSS SN~Ib/c template (SN~2006fo), and the Hsiao SN~Ia template.\footnote{Interestingly, despite its large uncertainty, the $K_s$ measurement has sufficient constraining power to disfavor a single SN~Ib/c template (that would be otherwise compatible with the bluer \HST data for an extinction in excess of 0.5 mag $E(B-V)$); we show the remaining SN~Ib/c light-curve template that is not disfavored in Figure~\ref{fig:lc}.} 

From the light curve fits we obtain relative likelihoods of $\LtypeT/\LtypeIa = \LmeanIaT$, $\Ltypebg/\LtypeIa = \LmeanIabg$, $\LtypeIbc/\LtypeIa = \LmeanIbc$, and $\LtypeII/\LtypeIa = \LmeanII$. The highest likelihood (best-fit) points in normal SN~Ia parameter space are slightly higher than for the SN~1991T-like template, but the average likelihood for the normal SNe~Ia population is lower due to points with extreme $x_1$ or $c$ values, i.e., those that do not match the data well. Because the likelihood for the SN~1991bg-like template is so remote, we do not consider this subtype further.

It is worth noting that the two CC SNe are $\gtrsim 2$ magnitudes fainter in absolute magnitude than any reasonable value for SN~SCP16C03 (we discuss the lensing amplification in Section~\ref{sec:lensinginterp}), so our typing analysis is very conservative since absolute magnitude is not included in the evaluation (see Appendix \ref{sec:obscuredCC}). However, we do not have enough templates derived from CC SNe in the same absolute magnitude range as SNe~Ia, so we include these fainter CC SNe for now. 

\subsection{The intrinsic fraction of SNe~Ia}

The next ingredient for computing an overall probability of being a SN~Ia is an estimate of the relative rates of CC SNe and SNe~Ia in the host galaxy. We follow \citet{meyers12} and estimate the {\it intrinsic} SN~Ia fraction, i.e., without imposing any detection limit. Starting from relations for the relative rates of SNe~Ia\footnote{Here, we use the ``prompt-and-delayed'' model, in which the SN~Ia rate is modeled as a linear combination of stellar mass and star-formation rate \citep{scannapieco05, mannucci06}. We use coefficients from the \citet{sullivan06} analysis. We note that the \citet{sullivan06} coefficients are applied to the SFR averaged over the previous 500 Myr, rather than the instantaneous SFR. As the SFR was higher in the past for this galaxy, our procedure conservatively underestimates the SN Ia rate associated with star formation. We note that the SN Ia rate is dominated by the mass component for either SFR averaging time period, so our results are insensitive to this choice. In any case, this is a simple approximation to the true delay-time distribution, which is close to $t^{-1}$ (see the review of \citealt{maoz12}). As an alternative, we also use the host-galaxy age and stellar mass with a $t^{-1}$ delay-time distribution (normalized as in \citealt{rodney14}), finding a somewhat higher SN~Ia rate. We use the prompt-and-delayed rate to be conservative.} 
and CC SNe\footnote{Assuming a Salpeter IMF \citep{salpeter55} with all stars from 8 to 40~M$_{\odot}$ forming CC SNe, there will be $6.8\times10^{-3}$ CC SNe per M$_{\odot}$ of formed stars. This assumption is a reasonable match to observed CC rates and the cosmic star-formation history \citep{madau14, strolger15}.}, 
based on SFR and stellar mass, along with our measurements of these properties --- $\log_{10}{(\textrm{M/M}_{\odot})} = \fastmass$ and $\dot{M} = \FASTSFR$~M$_{\odot}$/yr from Table~\ref{tab:hostprop2} of Section~\ref{sec:host} --- our estimated fraction of SNe~Ia is

\begin{eqnarray}
\fIa & =& \left[1 + \frac{R_{\mathrm{CC}}}{R_{\mathrm{\, Ia}}} \right]^{-1} \nonumber \\
 & = & \left[1 + \frac{6.8\times10^{-3}\ \mathrm{\dot{M}}}{5.3 \times 10^{-14}\ \mathrm{M} + 3.9 \times 10^{-4}\ \mathrm{\dot{M}}} \right]^{-1} \label{eq:rate} \\
 & = & \left[1 + \frac{\hostCCRateRest \ \mathrm{yr}^{-1}}{\hostIaRateRest \ \mathrm{yr}^{-1}} \right]^{-1} = \IaCombinedFractionHost\;  \nonumber
\end{eqnarray}

Thus, we expect that \IaCombinedPercentHost of the SNe exploding in this host galaxy at the epoch when SN~SCP16C03 was observed will be SNe~Ia. We note that \fIa as calculated here is essentially independent of the assumed amplification, as both the SFR and stellar mass have the same scaling with amplification and thus it cancels out of Equation~\ref{eq:rate}. Moreover, these rates integrate over all SNe, regardless of their luminosity, and amplification can not change the intrinsic ratio. We further note that \fIa\ does not depend on the spatial distribution of star formation in the host galaxy. For example, if there is \CombinedSFR~M$_{\odot}$/yr of star formation occurring in a satellite galaxy, then \IaCombinedPercentHost of the SNe in the combined system will still be SNe~Ia. Because all SNe are included, and because most CC SNe are fainter than most SNe~Ia, this calculation of \fIa\ will be lower than the actual observed SN~Ia fraction.

In Appendix~\ref{sec:obscuredCC} we perform an analysis in which typical SN luminosity distributions, and the detection efficiency of our survey, are included, as well as allowing for an extra starburst component added to the old stellar population from the best-fit \texttt{FAST} model. We address whether enhanced star-formation obscured by dust at the SN location, in combination with amplification by MOO~J1014+0038, could enhance the observed fraction of CC~SNe, concluding that for realistic amounts of amplification, it does not.

Since photometric typing is the primary means by which large upcoming SN cosmology surveys plan to obtain large numbers of SNe~Ia, it is worth noting that $f_{\mathrm{Ia}}$ could go as low as 0.05 for an extreme starburst in a low-mass host galaxy. Thus, this prior can significantly alter photometric classification probabilities across the full range of SN~Ia host galaxies.

\subsection{The intrinsic fraction of subtypes within each major type}

The final ingredient required for photometric classification of SN~SCP16C03 is the relative incidence in nature of the subtypes --- normal SN~Ia and SN~1991T-like subtypes among SNe~Ia, and SN~II and SN~Ibc among CC SNe --- corresponding to the light curve template categories that we use. We take a ratio of 3-1 for SNe~II to SNe~Ib/c \citep{li11}. We therefore apportion the CC fraction, $(1-\fIa)$, into $0.75\, (1-\fIa)$ for SNe~II and $0.25\, (1-\fIa)$ for SNe~Ibc. In volume-limited nearby samples, 3\% of SNe~Ia are pure SN~1991T-like \citep{scalzo12,silverman12}. Another 5\% of nearby SNe are SN~1999aa-like \citep{silverman12} while at high redshift we derive a SN~1999aa-like fraction of 1--3\% from \citet{balland09}. However, SN~1999aa-like SNe are only marginally overluminous, and so these can be treated with the other normal SNe. We therefore apportion $\fIa$ into $0.03\, \fIa$ for SN~1991T-like and $0.97\,\fIa$ for normal SNe~Ia (i.e., also excluding SN~1991bg-like or SN~2002cx-like SNe~Ia). Although these subtype fractions are derived at low redshift and may be different at $z=\shortredshift$, the assumed values do not strongly affect the classification as a SN~Ia. Thus, our estimated fractions of each type of SN, $f_{\type}$, are: $f_{\Ia\,\mathrm{Normal}}=\IaPercentHost$ for normal SN~Ia, $f_{\IaT}=\IaTPercentHost$ SN~1991T-like SN~Ia, $f_{\II}=\IIPercentHost$ SN~II, and $f_{\Ibc}=\IbcPercentHost$ SN~Ib/c.\footnote{The fractions here add up to $99\%$ because they are rounded, but the final probability is precisely normalized in Equation~\ref{eq:type}.}

\subsection{SN classification results}

Combining the likelihoods from the light-curve fitting, with the product of the fraction of SN types in nature and the SN~Ia fraction from the host-galaxy properties, we obtain our estimates of the overall probability of each type. Equation~\ref{eq:type} gives a \finalIaNprob chance that SN~SCP16C03 is a normal SN~Ia, a \finalIaTprob chance it is a SN~1991T-like SN~Ia, a \finalIbcprob chance it is a SN~Ib/c, and a \finalIIprob chance of it is a SN~II. This analysis is summarized in the top left panel of Figure~\ref{fig:lc}, which shows the highest (best-fit) relative likelihood for each template. The total probability that SN~SCP16C03 is a SN~Ia is $P(\Ia) + P(\IaT) = \finalIaAllprob$. 
As a final (unlikely) possibility, in Appendix~\ref{sec:proj} we evaluate the probability that the SN is merely projected onto the nominal host galaxy, rather than being hosted by it, and find it to be negligible.

\section{Light-Curve Parameters and Amplification Measurement} \label{sec:lcfit}

To compute an amplification estimate, and locate SN~SCP16C03 in the lower-redshift light-curve parameter distributions, we fit the photometry in Table~\ref{tab:photometry} with the SALT2 light-curve fitter. SALT2 fits a rest-frame model to the observer-frame photometry, extracting a date-of-maximum, amplitude (scaled by the inverse squared luminosity distance and typically quoted using a rest-frame $B$-band magnitude, $m_B$), light-curve shape ($x_1$), and color ($c$). We find $t_{\mathrm{max}} = 57474.6 \pm 1.4$~MJD, $m_B = 26.067\pm 0.041$~mag, $x_1=0.54\pm0.60$, and $c=0.039\pm0.051$. Our light-curve fit parameters are consistent with the center of the low-redshift $x_1$ and $c$ distributions, disfavoring strong evolution in the population mean of these parameters.\footnote{In order to quantify this, we have to subtract the measurement uncertainties. When we do this assuming a split-normal distribution (\url{https://github.com/rubind/salt2params}), we find that SN~SCP16C03 is at the $64\pm20$th percentile in $x_1$, and the $67^{+16}_{-22}$th percentile in $c$. If the (mild) evidence for SN light in the epoch before the detection is real, the rise-time is $\sim 20$ days in rest-frame $B$-band. This is slower than average for a SN~Ia, but still within the observed distribution \citep{hayden10}.} This test would have been difficult to perform with unlensed SNe in this redshift range due to selection effects (the 50\% completeness limit at this redshift for unlensed SNe is about $-19.1$, similar to the median absolute magnitude of SNe~Ia; \citealt{rodney15b}, \citealt{hayden16}) and the difficulty of photometrically typing SNe $\sim 1$ magnitude fainter than the amplified SN~SCP16C03.

In many previous analyses, linear shape and color standardization has been employed \citep{tripp98}, i.e.,
\begin{equation}
\mu_B = m_B + \alpha\ x_1 - \beta\ c - M_B
\end{equation}
where $\alpha$ and $\beta$ are the slope of the shape and color standardization relations, and $M_B$ is the absolute $B$-band magnitude. 
Here we assume $H_0=70$~km~s$^{-1}$~Mpc$^{-1}$, but it drops out of the final analysis of the amplification. 

We take our values of the standardization coefficients, $\alpha = 0.141, \beta = -3.101$, from \citet{betoule14}. Evidence is steadily increasing that the $x_1$ and $c$ relations are not linear; at a minimum, broken-linear relations are more accurate \citep{amanullah10, suzuki12, scolnic14, rubin15, scolnic16, mandel16}. For simplicity, we use linear relations here, as this SN is close to the center of both the $x_1$ and $c$ distributions, so applying non-linear relations makes little difference. The observed absolute magnitude has a weak dependence on host-galaxy stellar mass \citep{kelly10, sullivan10}, but this relation may be partially driven by progenitor age, and thus would weaken with redshift as young progenitors will dominate galaxies of both high and low stellar mass \citep{rigault13, childress14}. However, the lack of detectable emission lines in the host of SN~SCP16C03 decreases the likelihood that it arose from a young progenitor system. Although most of the likelihood for our galaxy is in the high-mass category ($>10^{10}$~M$_{\odot}$), we split the difference in $M_B$, taking the average of the two host-mass categories (for a value of $-19.085$), and allocating half of the difference as uncertainty (0.035 magnitudes). We assume 0.12 magnitudes of ``intrinsic'' dispersion in the absolute magnitude of high-redshift SNe~Ia \citep{rubin16b}. We also consider the various WFC3 IR systematic uncertainties highlighted in \citet{nordin14}: count-rate nonlinearity, flux-dependent PSF, SED-dependent PSF, and the uncertainty in the CALSPEC system \citep{bohlin07, bohlin14}. We evaluate the impact of each systematic uncertainty on the distance modulus by varying each one and refitting the SN distance. We then include these differences in the quadrature sum for the total distance uncertainty.

In order to complete the amplification measurement, we need an estimate of the distance modulus (in the absence of lensing). One method for obtaining this is to compare against a sample of unlensed SNe at similar redshifts \citep{patel14, rodney15}, but no such sample exists for SN~SCP16C03. We thus compute the estimated amplification by comparing the SN distance modulus estimate with a cosmological model. We take $\Omega_m=0.3089$ from \citet{planck15} (including all cosmological datasets), and for simplicity we assume a flat $\Lambda$CDM cosmology, but we use a large $\pm 0.03$ Gaussian uncertainty on $\Omega_m$. This yields a predicted distance modulus of $\mu=46.219 \pm 0.057$, again for $H_0=70$~km~s$^{-1}$~Mpc$^{-1}$ (as noted above, $H_0$ drops out of the analysis). As the uncertainty on our measured distance modulus is much larger than the uncertainty on this prediction, the impact of our assumptions is minor. We obtain a distance modulus estimate of \finaldistmod, and an amplification estimate of \finalmagnification, with almost all of the measurement uncertainty (0.22 mag) statistical, and not from the cosmological model or systematic uncertainties.

\section{Amplification Interpretation} \label{sec:lensinginterp}

As noted in \citet{nordin14}, we can either test a lensing-derived model with the measured supernova amplification, or improve the lensing-derived model by including our amplification measurement. For the moment, we choose the former, comparing against the weak-lensing (WL) measurements of \citet{kim17}, which are derived independently from this analysis. That analysis is based on galaxy shear measurements from the $F105W$ and $F140W$ imaging data; the total exposure time of each filter exceeded 16,000 s at the time of the analysis. The PSFs are modeled separately using globular cluster observations originally planned for the WFC3 on-orbit calibration. We measure a galaxy shear by fitting a PSF-convolved elliptical Gaussian profile to a galaxy image. After verifying the consistency of the results from both filters, we optimally combine the two shear measurements and create a single source catalog. The redshift distribution of the source population is estimated by utilizing the publicly available UVUDF photometric redshift catalog \citep{rafelski15}. Because of the limited angular scale, we assume a mass-concentration relation of \citet{duffy08} in our mass estimation of the cluster while fitting an NFW profile to tangential shears.

Given that the contours of the WL shear map appear to be reasonably smooth, we assume spherical symmetry for the cluster. We use Monte-Carlo methods to derive the predicted amplification at the location of the SN image, assuming Gaussian constraints on the centroid of the cluster and the virial mass, and find $0.61^{+0.20}_{-0.16}$ mag. This value is compatible at $1.6\, \sigma$ with the SN amplification measurement (assuming that the uncertainty on the difference is the quadrature sum of the amplification uncertainties). We note that a concentration of bright cluster galaxies lies slightly to the west of the lens center determined from weak lensing analysis.  Their proximity to the SN may boost its amplification. We show contours of the full map (from the median of the Monte-Carlo samples) in Figure~\ref{fig:postagestamp}. We have also used the amplification determined from the SN observations to try to estimate the concentration parameter, $c_{200}$. While at the current level of observation it is not possible to place strong constraints on the concentration, the inferred value for $c_{200}$ is consistent with N-body simulations based on $\Lambda \mathrm{CDM}$ at $z = 1.23$.

In previous work with high-redshift SNe, \citep[e.g.,][]{suzuki12, rubin13, nordin14}, we were able to make use of ``blinded'' analyses, in which the results are hidden until the analysis is finalized \citep[e.g.,][]{conley06, maccoun15}. In those works, we were able to blind both the final photometry and the final typing, as both improved over the course of the analysis. In our analysis of SN~SCP16C03, we approached the photometry and typing with a mature pipeline, and made the decision to trigger followup based on this pipeline. Those results cannot be taken to be blinded, although we know of no biases that were introduced. By contrast, the lensing comparison was only conducted after the rest of the analysis was frozen and is thus fully blinded.

\section{Summary}

We present our discovery and measurements of a lensed SN with a host-galaxy redshift of \shortredshift. The light curve of this SN most closely matches a SN~Ia, in both shape and color; there are only two known core collapse templates within a factor of 20 of the best fit SN~Ia likelihood (and with both of these requiring an additional two magnitudes of amplification to match the absolute magnitude of SNe~Ia).
We determine a robust host galaxy redshift of \redshift $\pm 0.0002$, from a VLT X-shooter spectrum displaying multiple absorption features. Measurements of the spectrum, SED fitting of the spectrum and broadband photometry (from the rest-frame UV to $J$ band), and measurement of the surface brightness profile consistently demonstrate a massive, low-SFR host galaxy. When combined with observational constraints on SN rates, the low-SFR environment implies that most of the SNe in this galaxy are expected to be SNe~Ia. When considering the likelihoods from the light curve fits to all SN subtypes, this leads to a \finalIaAllprob probability that this exceptional event is a SN~Ia. This makes it the highest redshift SN~Ia with a spectroscopic redshift and the highest redshift lensed SN of any type. Using the conventional SN~Ia standardization relation and assuming a \citet{planck15} cosmology, we estimate the SN amplification is \finalmagnificationflux in flux (1.1 mag), making this the most amplified SN~Ia discovered behind a galaxy cluster to date. We estimate only a 10\% chance of finding a SN~Ia at or beyond this redshift (using the high-redshift SN rate model in \citet{rodney14} up to $z=3$) at least this close to the center of one of our clusters (thus allowing it to be lensed), making this an exceptional discovery. The light-curve parameters of SN~SCP16C03 are close to the mode of the population distribution of lower redshift SNe, indicating that our first unbiased look at the $z \sim 2$ SN~Ia population shows no evidence of strong population drift in these parameters. We also find consistency between the estimated amplification from the weak-lensing-derived map and our Hubble diagram residual. Based on our work, it appears that there were normal SNe~Ia $\sim 11$ Gyr before the present.

\acknowledgements
Based on observations made with the NASA/ESA \Hubble, obtained at the Space Telescope Science Institute, which is operated by the Association of Universities for Research in Astronomy, Inc., under NASA contract NAS 5-26555. These observations are associated with programs GO-13677 and GO-14327. Additionally based on observations collected at the European Organisation for Astronomical Research in the Southern Hemisphere under ESO program IDs 096.A-0926 and 296.A-5051. This publication makes use of data products from the Two Micron All Sky Survey, which is a joint project of the University of Massachusetts and the Infrared Processing and Analysis Center/California Institute of Technology, funded by the National Aeronautics and Space Administration and the National Science Foundation. Based on observations made with the Gran Telescopio Canarias (GTC), installed at the Spanish Observatorio del Roque de los Muchachos of the Instituto de Astrofísica de Canarias, on the island of La Palma. Additional support was provided from NASA ADAP NNH16AC25I. Support for programs GO-13677, GO-14327, and SNAP-14163 were provided by NASA through a grant from the Space Telescope Science Institute, which is operated by the Association of Universities for Research in Astronomy, Inc., under NASA contract NAS 5-26555. Additional support was provided by the Director, Office of Science, Office of High Energy Physics, of the U.S. Department of Energy under contract No. DE-AC02-05CH11231. M. J. Jee acknowledges support from KASI and NRF of Korea to CGER. The work of P.E. and D.S. was carried out at Jet Propulsion Laboratory, California Institute of Technology, under a contract with NASA. We thank Rachel Bezanson for useful feedback on our assessment of the host-galaxy properties. We also thank Dan Maoz for a discussion of the WFC3 pipeline units. Finally, we acknowledge the anonymous referee, whose feedback greatly improved this manuscript.

\facilities{HST (WFC3), VLT:Kueyen, VLT:Yepun, GTC}

\appendix

\section{Could Dust and Amplification Conspire to Enhance the Core-Collapse Fraction?}\label{sec:obscuredCC}

Without photometry that covers the rest-frame far-IR, it is difficult to entirely rule out obscured star-formation from the host galaxy alone. If the star-formation were obscured but undetected in our \texttt{FAST} fit, and the resulting core-collapse supernovae were still visible, our calculation of $f_{\mathrm{Ia}}$ from Equation~\ref{eq:rate} would overestimate the SN~Ia fraction and thereby increase the estimated SN~Ia probability. In this section, we address this scenario by evaluating the amount of star-formation that could be hidden by dust, modifying Equation~\ref{eq:rate} to account for a hidden starburst, and then applying the See-Change apparent magnitude detection efficiency curve to the modified rates for each SN type. We find that a hidden starburst as the source of this SN is very unlikely and does not alter its classification as a SN~Ia.

We modify Equation~\ref{eq:rate} to include mass and SFR estimates from the \texttt{FAST} SED fit and a separate starburst component. We also modify the core-collapse and Ia rates by the fraction of supernovae that are detectable by our survey, estimated on a grid of amplification and the starburst dust values, $A_{V,SB}$. Both of these procedures are described below. Starting from Equation~\ref{eq:rate}, we parameterize the rates components as
\begin{eqnarray}
R_{\mathrm{CC}} & \equiv & \mathrm{A}\,\mathrm{\dot{M}_\mathrm{T}}\,\mathrm{D}_{\mathrm{CC}} \label{eqn:appArates}\\
R_{\mathrm{Ia}} & \equiv & \left(\mathrm{B}\,\mathrm{M}_\mathrm{T}+\mathrm{C}\,\dot{\mathrm{M}}_\mathrm{T}\right)\,\mathrm{D}_{\mathrm{Ia}} \nonumber
\end{eqnarray}
with A$=6.8\times10^{-3}$, B$=5.3\times10^{-14}$, C$=3.9\times10^{-4}$, $\dot{\mathrm{M}}_\mathrm{T}\equiv\left(\dot{\mathrm{M}}_{\mathrm{FAST}}+\dot{\mathrm{M}}_{\mathrm{SB}}\right)$, $\mathrm{M}_\mathrm{T}\equiv\left(\mathrm{M}_{\mathrm{FAST}}+\mathrm{M}_{\mathrm{SB}}\right)$, and D$_{\textrm{CC}}$ and D$_{\textrm{Ia}}$ the fractions of the CC and Ia absolute magnitude distributions that are visible in our survey. As shown below, these fractions are estimated on a grid of amplification and $A_{V,SB}$ values. Since M and $\dot{\mathrm{M}}$ scale linearly with amplification, the detection fractions are the only components of this modified $f_{\mathrm{Ia}}$ equation that scale with amplification. The amount of starburst $\mathrm{M}_\mathrm{SB}$ and $\dot{\mathrm{M}_\mathrm{SB}}$ allowed by the host galaxy photometry and spectroscopy, along with these detection fractions, are the critical components for investigating whether dusty star-formation could bias our typing procedure towards high SN Ia rates.

To determine the amount of star-formation that could be hidden in a dusty starburst component in the galaxy SED, we evaluate the $\chi^2$ value of the best-fit  \texttt{FAST} template with a starburst component added, on a grid of $A_{V,SB}$ and starburst flux values. For low starburst fluxes, this component primarily affects the GTC $g$-band, but at higher fluxes, it can affect the GTC $r$ and $i$ bands, the $HST$ WFC3 UVIS $F814W$, and the bluest regions of the X-shooter spectrum. This provides a relatively tight constraint on the strength of a starburst component added to the SED (see i.e. Figure~\ref{fig:SED}, where the GTC $r$ and $i$ and \HST $F814W$ are already slightly over-estimated by the \texttt{FAST} SED fit); at $A_{V,SB}=0$, $\dot{\mathrm{M}}_{\mathrm{SB}}=0.07$ M$_\odot$/yr, while at $A_{V,SB}=3.0$, $\dot{\mathrm{M}}_{\mathrm{SB}}=1.09$ M$_\odot$/yr. The $\Delta \chi^2$ improvement is only $\sim0.05$ for the additional 2 degrees of freedom, meaning that the simpler, starburst-free model is preferred statistically. Still, in Equation \ref{eqn:appArates}, we will include this best-fit $\dot{\mathrm{M}}_{\mathrm{SB}}$ in all SFR estimates for determination of $f_{\mathrm{Ia}}$ in the host galaxy.

\begin{figure*}
\begin{centering}
\includegraphics[width= 0.9\textwidth]{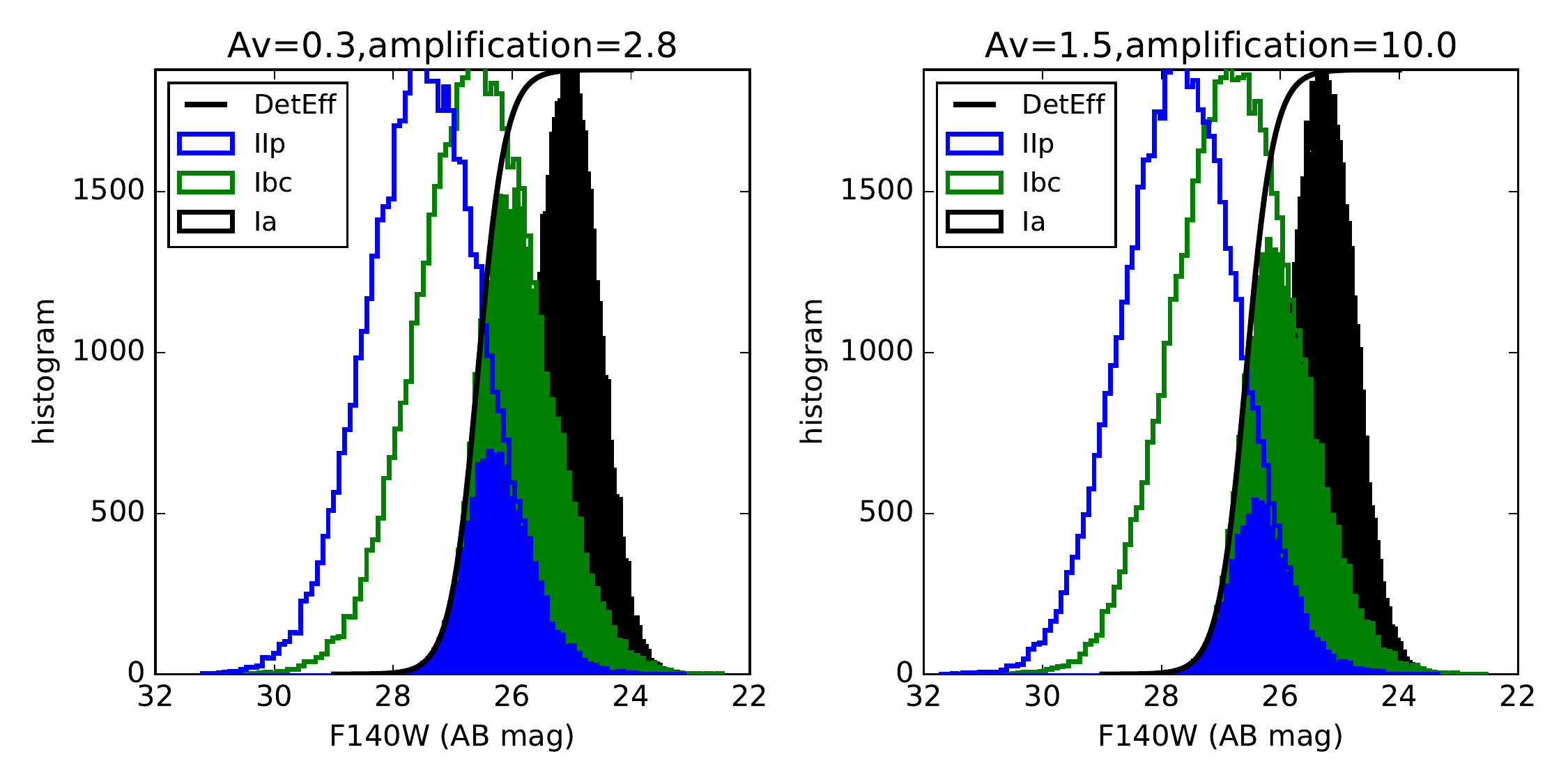}
\caption{Monte Carlo simulation of supernova detectability by type, as the amount of dust is increased, and accounting for the See Change detection efficiency versus magnitude. The ratio of the area under the filled histograms to the open histograms is D$_{\textrm{CC}}$ or D$_{\textrm{Ia}}$ in Equation~\ref{eqn:appArates} (for D$_{\textrm{CC}}$, the IIp and Ibc SNe are combined with equal weight). We find that for reasonable amplification values, the observable CC SN rate is highly impacted by the truncation of the detection efficiency curve, leading the observable SN rate in this galaxy to move towards SNe Ia at higher starburst fractions (see Figure~\ref{fig:fIa_starburst}).}
\label{fig:Av_sim}
\end{centering}
\end{figure*}

To determine the detectable fractions of SNe of each type, we perform a simple Monte Carlo using the absolute magnitude distributions from \citet{richardson14}, over a grid of $A_{V,SB}$ and amplification values. For each grid point, we draw absolute magnitudes from a normal distribution for each SN type given by \citet{richardson14}. Using \texttt{sncosmo}, we build the spectral time-series for each type using SN~2007md for Type IIp, SN~2006fo for Type~Ibc\footnote{These two supernovae are the only core-collapse light curves with fit probabilities within a factor of 20 of the best Ia template in Section \ref{sec:photclass}}, and the Hsiao template \citep{hsiao07} for Type~Ia. We apply the given $A_{V,SB}$ to the spectral time-series and calculate its apparent magnitude at maximum brightness in the $F140W$ filter at $z=2.2216$ assuming a \citet{planck15} cosmology (see Figure~\ref{fig:Av_sim}).

During the execution of the See Change survey, we planted blinded fake supernovae, and only unblinded after we had determined whether all objects were valuable enough to trigger $HST$ follow-up. As a result, we have a robust measurement of our detection efficiency curve in apparent magnitude (see black line in Figure~\ref{fig:Av_sim}). For each simulated supernova in our Monte Carlo simulation, we determine its likelihood of detection using this curve, so for each grid point in $A_{V,SB}$ and amplification, this provides\footnote{In effect, this is a numerical integration of the multiplication of our detection efficiency curve with the apparent magnitude distribution after accounting for dust and the assumed \citet{planck15} cosmology.} D$_{\textrm{CC}}$ and D$_{\textrm{Ia}}$ for use in Equation~\ref{eqn:appArates}.

\begin{figure*}
\begin{centering}
\includegraphics[width= 0.7\textwidth]{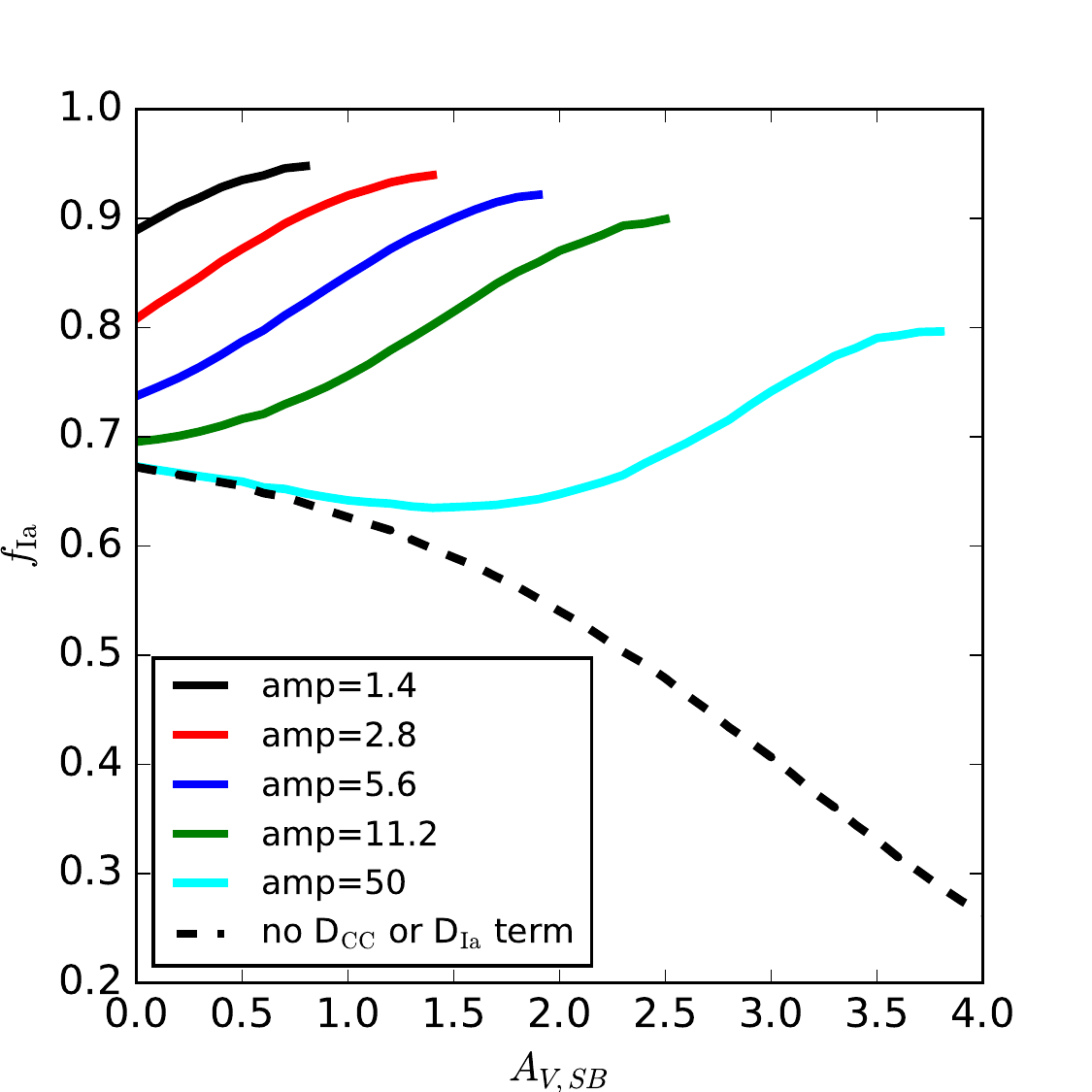}
\caption{$f_{\mathrm{Ia}}$ using the modified rates from Equation~\ref{eqn:appArates}, as a function of $A_{V,SB}$, for various amplification values with (\textit{solid}) and without (\textit{dashed}) the detection efficiency term. The $f_{\mathrm{Ia}}$ curves for different amplifications are shown only for D$_{\textrm{CC}}>0.05$. For reasonable values of amplification, the CC SN magnitude distribution is truncated by the detection efficiency, and $R_{\mathrm{CC}}$ is dominated by the quickly declining D$_{\textrm{CC}}$ term. For extremely high amplifications, the entire CC distribution is visible at low $A_{V,SB}$, so D$_{\textrm{CC}}=1$; in this scenario, a small enhancement to $R_{\mathrm{CC}}$ occurs as the hidden starburst component increases, causing $f_{\mathrm{Ia}}$ to briefly decline with $A_{V,SB}$ until D$_{\textrm{CC}}$ declines. We conclude that our primary typing algorithm, Equation~\ref{eq:type}, using $f_{\mathrm{Ia}}$ as in Equation~\ref{eq:rate}, is conservative for the SN Ia probability, since the observable $R_{\mathrm{CC}}$ is in reality lower due to our magnitude limits. }
\label{fig:fIa_starburst}
\end{centering}
\end{figure*}

Figure~\ref{fig:fIa_starburst} shows the result for $f_{\mathrm{Ia}}$ after accounting for the starburst component and the detection efficiency of our survey. We find that for reasonable amplification values, the core-collapse rate is strongly truncated by the detection limit term in Equation~\ref{eqn:appArates}; the initial detection efficiency for CC SNe, D$_{\textrm{CC}}$ at $A_{V,SB}=0$, does not cross $95\%$ until an amplification of $\sim17$. As a result, $f_{\mathrm{Ia}}$ increases with increasing dusty star-formation, and when combined with the fact that even at low $A_{V,SB}$ the CC distributions are truncated (see i.e. Figure~\ref{fig:Av_sim}), this implies that our main typing algorithm underestimates the SN Ia probability by not including absolute magnitude.

\section{Is the redshift $\shortredshift$ galaxy the SN host?}\label{sec:proj}

For use in Section \ref{sec:class}, we now estimate the probability that this SN is merely projected on the $z=\shortredshift$ galaxy, rather than hosted by it. As in Section~\ref{sec:class}, we consider the average likelihoods and rates for each type of SNe; for projected SNe, we must also consider these quantities over a range in redshift. We thus need three new ingredients 1) The likelihood as a function of redshift for CC SNe and SNe~Ia. 2) The relative rates (per unit area, at the SN location) in the field and in the host. 3) The fraction of stellar mass and star-formation (and thus rate) that is excluded for this SN, as we do not see a second galaxy on top of the presumptive host.

For the likelihood as a function of redshift, we run the photometric typing at a range of redshifts and compute the equivalent redshift width for each type. This is defined relative to the likelihood at $z=\shortredshift$, in the sense that averaging the likelihood over all redshifts gives the same answer as multiplying the $z=\shortredshift$ value by the equivalent redshift width. We find this quantity is \DzIa for normal SNe~Ia, \DzIaT for SN~1991T-like SNe~Ia, \DzIbc for SNe~Ib/c, and \DzII for SNe~II. The CC SN values are larger than for SNe~Ia. The CC SNe have more heterogeneity, so they are consistent with the light curve over a wider range in redshift (i.e., a SN-only photo-z is not as accurate for a CC SN). As a special case, to test whether this SN could be an intra-cluster SN within MOO~J1014+0038 having a chance alignment with a background galaxy, we also run the typing at the cluster redshift of 1.23. The best-fit likelihood values are \clusterLL times lower, strongly disfavoring this possibility.

Now we need to consider the relative rates compared to the $z=\shortredshift$ galaxy. From \citet{rodney14}, the volumetric SN~Ia rate in this redshift range is around $0.5 \times 10^{-4}$ yr$^{-1}$ Mpc$^{-3}$ $h^3_{70}$. From \citet{strolger15}, the volumetric CC SN rate in this redshift range is around $3.5 \times 10^{-4}$ yr$^{-1}$ Mpc$^{-3}$ $h^3_{70}$. We must now transform these rates (and the above rates in the host galaxy) to observer-frame-rate surface densities evaluated at the SN location. For convenience, we work in per square arcsec units (although this is an arbitrary choice). From the factors in Equation~\ref{eq:rate}, we estimate the global rates in the galaxy (\hostIaRateRest for SNe~Ia and \hostCCRateRest for CC SNe, both per rest-frame year); assuming the SN rates in the galaxy are roughly proportional to the rest-frame $B$-band ($F140W$) surface brightness \citep{heringer17}, we can find the rate surface density. Using a small aperture at the location of the SN (on an $F140W$ stacked image with no SN light), we measure the surface brightness. Multiplying this surface brightness by an area of one square arcsecond corresponds to 10\% the luminosity of the $z=\shortredshift$ galaxy. Thus, the SN rate per square arcsec is about 10\% of the global galaxy rate, or \hostArcSecCCRateRest for CC SNe and \hostArcSecIaRateRest for SNe~Ia (both per rest-frame year). For the volumetric rates, we would naively find \volumeArcsecCCRateRestNaive CC SNe and \volumeArcsecIaRateRestNaive SNe~Ia per square arcsec, per unit redshift, per rest-frame year, but we must lower these by the amplification (as the volume behind the cluster is lowered by the same factor). To be conservative, we use the (smaller) weak-lensing amplification estimate of \weaklensingonlyamplflux, for rates of \volumeArcsecCCRateRestDelensed (CC SNe) and \volumeArcsecIaRateRestDelensed (SNe~Ia).

In some sense, the above volumetric rates count all expected galaxies along the line of sight, and thus double count the $z=\shortredshift$ galaxy. They also count galaxies that would be clearly visible superposed on the $z=\shortredshift$ galaxy (the vast majority of SN hosts). Thus, we must scale the rates by the fraction of star formation and stellar mass in galaxies too faint to detect in projection. Using only SN-free $F140W$ images, we measure a 95\% upper limit for a point-like galaxy\footnote{The FWHM of the PSF is 1 kpc at this redshift. Assuming a larger galaxy will give a somewhat weaker flux limit, but this has little quantitative effect on our conclusions. Assuming a magnitude limit one magnitude brighter only affects the projection probability by 0.1 percentage points.} at the SN location of 0.2 e$^-$/s (after subtracting a smooth galaxy model for the $z=\shortredshift$ galaxy). This is 28.2 (AB) as observed, or $\gtrsim 28.8$ after de-lensing. Assuming a redshift $\sim \shortredshift$, this corresponds to an absolute magnitude fainter than $16.2$ in rest-frame $B$ band. Comparing to the rest-frame $B$-band luminosity function at this redshift \citep{marchesini07}, we see that our limit is almost six magnitudes fainter than $M^*$. For a faint-end power law slope of $-1.5$, this would imply that 93\% of $B$-band luminosity is excluded based on our upper limit. We apply this same factor to the SN rates; as we show, the probability of projection is small enough that reasonable variation in these assumptions does not affect the conclusion.

Finally, we complete the estimation of the projection probability, shown in Table~\ref{tab:exttyping}. We collect the relative likelihood values from Section~\ref{sec:class} and put them in the third column. The rates (per unit area and time for the $z=\shortredshift$ galaxy, and per unit area, time, and $\Delta z$ for projections) are summarized in the fourth column. These are the only two quantities needed in the case that the $z=\shortredshift$ galaxy is the host. However, we multiply by the redshift equivalent width for the projections, giving more probability to types of SNe which are consistent with the light curve over a longer redshift baseline. In the fifth column, we take the product of these factors, giving the rate-normalized likelihood for each type of SNe. In the final column, we show these values after normalizing to one. Summing the probabilities in the last four rows, we find the probability of a projection is $\sim 2\times 10^{-3}$.

\begin{deluxetable}{ccccc|cc}[h]
\caption{Extended typing analysis. The final column gives the probability of each scenario.\label{tab:exttyping}}
\tablehead{\colhead{Host} & \colhead{SN Type} & \colhead{Relative Likelihood} & \colhead{Rate (Square Arcsec)$^{-1}$ yr$^{-1}$} & \colhead{Equivalent Redshift Width} &  \colhead{Product} &  \colhead{Normalized} \\
\colhead{} & \colhead{} & \colhead{} & \colhead{($\Delta z ^{-1}$ for projections)} & \colhead{} &  \colhead{} &  \colhead{}}
\startdata
$z=\shortredshift$ & Ia & \LmeanIa & \IaRateLocalHost & $\cdots$ &\IaRateHostUnnorm & \IaRateHostNorm \\
$z=\shortredshift$ & Ia 1991T & \LmeanIaT & \IaTRateLocalHost &$\cdots$ &\IaTRateHostUnnorm & \IaTRateHostNorm \\
$z=\shortredshift$ & Ib/c & \LmeanIbc & \IbcRateLocalHost &$\cdots$ &\IbcRateHostUnnorm & \IbcRateHostNorm \\
$z=\shortredshift$ & II & \LmeanII & \IIRateLocalHost &$\cdots$ &\IIRateHostUnnorm & \IIRateHostNorm \\
\hline
Projection & Ia & \LmeanIa & \IaRateLocalProj & \DzIa &\IaRateProjUnnorm & \IaRateProjNorm \\
Projection & Ia 1991T & \LmeanIaT & \IaTRateLocalProj & \DzIaT &\IaTRateProjUnnorm & \IaTRateProjNorm \\
Projection & Ib/c & \LmeanIbc & \IbcRateLocalProj & \DzIbc &\IbcRateProjUnnorm & \IbcRateProjNorm \\
Projection & II & \LmeanII & \IIRateLocalProj & \DzII &\IIRateProjUnnorm & \IIRateProjNorm \\
\enddata
\end{deluxetable}


\begin{thebibliography}{}
\expandafter\ifx\csname natexlab\endcsname\relax\def\natexlab#1{#1}\fi

\bibitem[{{Amanullah} {et~al.}(2010){Amanullah}, {Lidman}, {Rubin}, {Aldering},
  {Astier}, {Barbary}, {Burns}, {Conley}, {Dawson}, {Deustua}, {Doi}, {Fabbro},
  {Faccioli}, {Fakhouri}, {Folatelli}, {Fruchter}, {Furusawa}, {Garavini},
  {Goldhaber}, {Goobar}, {Groom}, {Hook}, {Howell}, {Kashikawa}, {Kim}, {Knop},
  {Kowalski}, {Linder}, {Meyers}, {Morokuma}, {Nobili}, {Nordin}, {Nugent},
  {{\"O}stman}, {Pain}, {Panagia}, {Perlmutter}, {Raux}, {Ruiz-Lapuente},
  {Spadafora}, {Strovink}, {Suzuki}, {Wang}, {Wood-Vasey}, {Yasuda}, \&
  {Supernova Cosmology Project}}]{amanullah10}
{Amanullah}, R., {Lidman}, C., {Rubin}, D., {et~al.} 2010, \apj, 716, 712

\bibitem[{{Amanullah} {et~al.}(2011){Amanullah}, {Goobar}, {Cl{\'e}ment},
  {Cuby}, {Dahle}, {Dahl{\'e}n}, {Hjorth}, {Fabbro}, {J{\"o}nsson}, {Kneib},
  {Lidman}, {Limousin}, {Milvang-Jensen}, {M{\"o}rtsell}, {Nordin}, {Paech},
  {Richard}, {Riehm}, {Stanishev}, \& {Watson}}]{amanullah11}
{Amanullah}, R., {Goobar}, A., {Cl{\'e}ment}, B., {et~al.} 2011, \apjl, 742, L7

\bibitem[{{Balland} {et~al.}(2009){Balland}, {Baumont}, {Basa}, {Mouchet},
  {Howell}, {Astier}, {Carlberg}, {Conley}, {Fouchez}, {Guy}, {Hardin}, {Hook},
  {Pain}, {Perrett}, {Pritchet}, {Regnault}, {Rich}, {Sullivan}, {Antilogus},
  {Arsenijevic}, {Le Du}, {Fabbro}, {Lidman}, {Mour{\~a}o},
  {Palanque-Delabrouille}, {P{\'e}contal}, \& {Ruhlmann-Kleider}}]{balland09}
{Balland}, C., {Baumont}, S., {Basa}, S., {et~al.} 2009, \aap, 507, 85

\bibitem[{{Barro} {et~al.}(2017){Barro}, {Faber}, {Koo}, {Dekel}, {Fang},
  {Trump}, {P{\'e}rez-Gonz{\'a}lez}, {Pacifici}, {Primack}, {Somerville},
  {Yan}, {Guo}, {Liu}, {Ceverino}, {Kocevski}, \& {McGrath}}]{barro17}
{Barro}, G., {Faber}, S.~M., {Koo}, D.~C., {et~al.} 2017, \apj, 840, 47

\bibitem[{{Belfiore} {et~al.}(2016){Belfiore}, {Maiolino}, {Maraston},
  {Emsellem}, {Bershady}, {Masters}, {Yan}, {Bizyaev}, {Boquien}, {Brownstein},
  {Bundy}, {Drory}, {Heckman}, {Law}, {Roman-Lopes}, {Pan}, {Stanghellini},
  {Thomas}, {Weijmans}, \& {Westfall}}]{belfiore16}
{Belfiore}, F., {Maiolino}, R., {Maraston}, C., {et~al.} 2016, \mnras, 461,
  3111

\bibitem[{{Bell} \& {de Jong}(2001)}]{bell01}
{Bell}, E.~F., \& {de Jong}, R.~S. 2001, \apj, 550, 212

\bibitem[{{Betoule} {et~al.}(2014){Betoule}, {Kessler}, {Guy}, {Mosher},
  {Hardin}, {Biswas}, {Astier}, {El-Hage}, {Konig}, {Kuhlmann}, {Marriner},
  {Pain}, {Regnault}, {Balland}, {Bassett}, {Brown}, {Campbell}, {Carlberg},
  {Cellier-Holzem}, {Cinabro}, {Conley}, {D'Andrea}, {DePoy}, {Doi}, {Ellis},
  {Fabbro}, {Filippenko}, {Foley}, {Frieman}, {Fouchez}, {Galbany}, {Goobar},
  {Gupta}, {Hill}, {Hlozek}, {Hogan}, {Hook}, {Howell}, {Jha}, {Le Guillou},
  {Leloudas}, {Lidman}, {Marshall}, {M{\"o}ller}, {Mour{\~a}o}, {Neveu},
  {Nichol}, {Olmstead}, {Palanque-Delabrouille}, {Perlmutter}, {Prieto},
  {Pritchet}, {Richmond}, {Riess}, {Ruhlmann-Kleider}, {Sako}, {Schahmaneche},
  {Schneider}, {Smith}, {Sollerman}, {Sullivan}, {Walton}, \&
  {Wheeler}}]{betoule14}
{Betoule}, M., {Kessler}, R., {Guy}, J., {et~al.} 2014, \aap, 568, A22

\bibitem[{{Bohlin}(2007)}]{bohlin07}
{Bohlin}, R.~C. 2007, in Astronomical Society of the Pacific Conference Series,
  Vol. 364, The Future of Photometric, Spectrophotometric and Polarimetric
  Standardization, ed. C.~{Sterken}, 315

\bibitem[{{Bohlin} {et~al.}(2014){Bohlin}, {Gordon}, \& {Tremblay}}]{bohlin14}
{Bohlin}, R.~C., {Gordon}, K.~D., \& {Tremblay}, P.-E. 2014, \pasp, 126, 711

\bibitem[{{Brammer} {et~al.}(2011){Brammer}, {Whitaker}, {van Dokkum},
  {Marchesini}, {Franx}, {Kriek}, {Labb{\'e}}, {Lee}, {Muzzin}, {Quadri},
  {Rudnick}, \& {Williams}}]{brammer11}
{Brammer}, G.~B., {Whitaker}, K.~E., {van Dokkum}, P.~G., {et~al.} 2011, \apj,
  739, 24

\bibitem[{{Brodwin} {et~al.}(2015){Brodwin}, {Greer}, {Leitch}, {Stanford},
  {Gonzalez}, {Gettings}, {Abdulla}, {Carlstrom}, {Decker}, {Eisenhardt},
  {Lin}, {Mantz}, {Marrone}, {McDonald}, {Stalder}, {Stern}, \&
  {Wylezalek}}]{brodwin15}
{Brodwin}, M., {Greer}, C.~H., {Leitch}, E.~M., {et~al.} 2015, \apj, 806, 26

\bibitem[{{Bruzual} \& {Charlot}(2003)}]{bruzual03}
{Bruzual}, G., \& {Charlot}, S. 2003, \mnras, 344, 1000

\bibitem[{{Calzetti} {et~al.}(2010){Calzetti}, {Wu}, {Hong}, {Kennicutt},
  {Lee}, {Dale}, {Engelbracht}, {van Zee}, {Draine}, {Hao}, {Gordon},
  {Moustakas}, {Murphy}, {Regan}, {Begum}, {Block}, {Dalcanton}, {Funes}, {Gil
  de Paz}, {Johnson}, {Sakai}, {Skillman}, {Walter}, {Weisz}, {Williams}, \&
  {Wu}}]{calzetti10}
{Calzetti}, D., {Wu}, S.-Y., {Hong}, S., {et~al.} 2010, \apj, 714, 1256

\bibitem[{{Cappellari}(2017)}]{ppxf17}
{Cappellari}, M. 2017, \mnras, 466, 798

\bibitem[{{Cappellari} \& {Emsellem}(2004)}]{ppxf04}
{Cappellari}, M., \& {Emsellem}, E. 2004, \pasp, 116, 138

\bibitem[{{Chabrier}(2003)}]{chabrier03}
{Chabrier}, G. 2003, \pasp, 115, 763

\bibitem[{{Charlot} \& {Fall}(2000)}]{charlot00}
{Charlot}, S., \& {Fall}, S.~M. 2000, \apj, 539, 718

\bibitem[{{Childress} {et~al.}(2014){Childress}, {Wolf}, \&
  {Zahid}}]{childress14}
{Childress}, M.~J., {Wolf}, C., \& {Zahid}, H.~J. 2014, \mnras, 445, 1898

\bibitem[{Clough {et~al.}(1992)Clough, Iacono, \& Moncet}]{clough92}
Clough, S.~A., Iacono, M.~J., \& Moncet, J.-L. 1992, Journal of Geophysical
  Research: Atmospheres, 97, 15761

\bibitem[{{Clough} {et~al.}(2005){Clough}, {Shephard}, {Mlawer}, {Delamere},
  {Iacono}, {Cady-Pereira}, {Boukabara}, \& {Brown}}]{clough05}
{Clough}, S.~A., {Shephard}, M.~W., {Mlawer}, E.~J., {et~al.} 2005, \jqsrt, 91,
  233

\bibitem[{{Conley} {et~al.}(2006){Conley}, {Goldhaber}, {Wang}, {Aldering},
  {Amanullah}, {Commins}, {Fadeyev}, {Folatelli}, {Garavini}, {Gibbons},
  {Goobar}, {Groom}, {Hook}, {Howell}, {Kim}, {Knop}, {Kowalski}, {Kuznetsova},
  {Lidman}, {Nobili}, {Nugent}, {Pain}, {Perlmutter}, {Smith}, {Spadafora},
  {Stanishev}, {Strovink}, {Thomas}, {Wood-Vasey}, \& {Supernova Cosmology
  Project}}]{conley06}
{Conley}, A., {Goldhaber}, G., {Wang}, L., {et~al.} 2006, \apj, 644, 1

\bibitem[{{Conroy} {et~al.}(2009){Conroy}, {Gunn}, \& {White}}]{con:fsps1}
{Conroy}, C., {Gunn}, J.~E., \& {White}, M. 2009, \apj, 699, 486

\bibitem[{{Davidzon} {et~al.}(2017){Davidzon}, {Ilbert}, {Laigle}, {Coupon},
  {McCracken}, {Delvecchio}, {Masters}, {Capak}, {Hsieh}, {Le F{\`e}vre},
  {Tresse}, {Bethermin}, {Chang}, {Faisst}, {Le Floc'h}, {Steinhardt}, {Toft},
  {Aussel}, {Dubois}, {Hasinger}, {Salvato}, {Sanders}, {Scoville}, \&
  {Silverman}}]{davidson17}
{Davidzon}, I., {Ilbert}, O., {Laigle}, C., {et~al.} 2017, \aap, 605, A70

\bibitem[{{Decker} {et~al.}(in prep.)}]{decker16}
{Decker}, B., {et~al.} in prep.

\bibitem[{{Dom{\'{\i}}nguez S{\'a}nchez} {et~al.}(2011){Dom{\'{\i}}nguez
  S{\'a}nchez}, {Pozzi}, {Gruppioni}, {Cimatti}, {Ilbert}, {Pozzetti},
  {McCracken}, {Capak}, {Le Floch}, {Salvato}, {Zamorani}, {Carollo},
  {Contini}, {Kneib}, {Le F{\`e}vre}, {Lilly}, {Mainieri}, {Renzini},
  {Scodeggio}, {Bardelli}, {Bolzonella}, {Bongiorno}, {Caputi}, {Coppa},
  {Cucciati}, {de la Torre}, {de Ravel}, {Franzetti}, {Garilli}, {Iovino},
  {Kampczyk}, {Knobel}, {Kova{\v c}}, {Lamareille}, {Le Borgne}, {Le Brun},
  {Maier}, {Mignoli}, {Pell{\'o}}, {Peng}, {Perez-Montero}, {Ricciardelli},
  {Silverman}, {Tanaka}, {Tasca}, {Tresse}, {Vergani}, \&
  {Zucca}}]{dominguez11}
{Dom{\'{\i}}nguez S{\'a}nchez}, H., {Pozzi}, F., {Gruppioni}, C., {et~al.}
  2011, \mnras, 417, 900

\bibitem[{{Duffy} {et~al.}(2008){Duffy}, {Schaye}, {Kay}, \& {Dalla
  Vecchia}}]{duffy08}
{Duffy}, A.~R., {Schaye}, J., {Kay}, S.~T., \& {Dalla Vecchia}, C. 2008,
  \mnras, 390, L64

\bibitem[{{Feulner} {et~al.}(2005){Feulner}, {Gabasch}, {Salvato}, {Drory},
  {Hopp}, \& {Bender}}]{feulner05}
{Feulner}, G., {Gabasch}, A., {Salvato}, M., {et~al.} 2005, \apjl, 633, L9

\bibitem[{{Freudling} {et~al.}(2013){Freudling}, {Romaniello}, {Bramich},
  {Ballester}, {Forchi}, {Garc{\'{\i}}a-Dabl{\'o}}, {Moehler}, \&
  {Neeser}}]{2013A&A...559A..96F}
{Freudling}, W., {Romaniello}, M., {Bramich}, D.~M., {et~al.} 2013, \aap, 559,
  A96

\bibitem[{{Gettings} {et~al.}(2012){Gettings}, {Gonzalez}, {Stanford},
  {Eisenhardt}, {Brodwin}, {Mancone}, {Stern}, {Zeimann}, {Masci}, {Papovich},
  {Tanaka}, \& {Wright}}]{gettings12}
{Gettings}, D.~P., {Gonzalez}, A.~H., {Stanford}, S.~A., {et~al.} 2012, \apjl,
  759, L23

\bibitem[{{Gilliland} {et~al.}(1999){Gilliland}, {Nugent}, \&
  {Phillips}}]{gilliland99}
{Gilliland}, R.~L., {Nugent}, P.~E., \& {Phillips}, M.~M. 1999, \apj, 521, 30

\bibitem[{{Gonzalez} {et~al.}(2015){Gonzalez}, {Decker}, {Brodwin},
  {Eisenhardt}, {Marrone}, {Stanford}, {Stern}, {Wylezalek}, {Aldering},
  {Abdulla}, {Boone}, {Carlstrom}, {Fagrelius}, {Gettings}, {Greer}, {Hayden},
  {Leitch}, {Lin}, {Mantz}, {Muchovej}, {Perlmutter}, \&
  {Zeimann}}]{gonzalez15}
{Gonzalez}, A.~H., {Decker}, B., {Brodwin}, M., {et~al.} 2015, \apjl, 812, L40

\bibitem[{{Gonzalez} {et~al.}(in prep.)}]{gonzalezInPrep}
{Gonzalez}, A.~H., {et~al.} in prep.

\bibitem[{{Goobar} {et~al.}(2009){Goobar}, {Paech}, {Stanishev}, {Amanullah},
  {Dahl{\'e}n}, {J{\"o}nsson}, {Kneib}, {Lidman}, {Limousin}, {M{\"o}rtsell},
  {Nobili}, {Richard}, {Riehm}, \& {von Strauss}}]{goobar09}
{Goobar}, A., {Paech}, K., {Stanishev}, V., {et~al.} 2009, \aap, 507, 71

\bibitem[{Goobar {et~al.}(2017)Goobar, Amanullah, Kulkarni, Nugent, Johansson,
  Steidel, Law, M{\"o}rtsell, Quimby, Blagorodnova, Brandeker, Cao, Cooray,
  Ferretti, Fremling, Hangard, Kasliwal, Kupfer, Lunnan, Masci, Miller,
  Nayyeri, Neill, Ofek, Papadogiannakis, Petrushevska, Ravi, Sollerman,
  Sullivan, Taddia, Walters, Wilson, Yan, \& Yaron}]{goobar16}
Goobar, A., Amanullah, R., Kulkarni, S.~R., {et~al.} 2017, Science, 356, 291

\bibitem[{{Gullikson} {et~al.}(2014){Gullikson}, {Dodson-Robinson}, \&
  {Kraus}}]{gullikson14}
{Gullikson}, K., {Dodson-Robinson}, S., \& {Kraus}, A. 2014, \aj, 148, 53

\bibitem[{{Guy} {et~al.}(2007){Guy}, {Astier}, {Baumont}, {Hardin}, {Pain},
  {Regnault}, {Basa}, {Carlberg}, {Conley}, {Fabbro}, {Fouchez}, {Hook},
  {Howell}, {Perrett}, {Pritchet}, {Rich}, {Sullivan}, {Antilogus}, {Aubourg},
  {Bazin}, {Bronder}, {Filiol}, {Palanque-Delabrouille}, {Ripoche}, \&
  {Ruhlmann-Kleider}}]{guy07}
{Guy}, J., {Astier}, P., {Baumont}, S., {et~al.} 2007, \aap, 466, 11

\bibitem[{{Hayden} {et~al.}(in prep.)}]{hayden16}
{Hayden}, B., {et~al.} in prep.

\bibitem[{{Hayden} {et~al.}(2010){Hayden}, {Garnavich}, {Kessler}, {Frieman},
  {Jha}, {Bassett}, {Cinabro}, {Dilday}, {Kasen}, {Marriner}, {Nichol},
  {Riess}, {Sako}, {Schneider}, {Smith}, \& {Sollerman}}]{hayden10}
{Hayden}, B.~T., {Garnavich}, P.~M., {Kessler}, R., {et~al.} 2010, \apj, 712,
  350

\bibitem[{{Heringer} {et~al.}(2017){Heringer}, {Pritchet}, {Kezwer}, {Graham},
  {Sand}, \& {Bildfell}}]{heringer17}
{Heringer}, E., {Pritchet}, C., {Kezwer}, J., {et~al.} 2017, \apj, 834, 15

\bibitem[{{Hsiao} {et~al.}(2007){Hsiao}, {Conley}, {Howell}, {Sullivan},
  {Pritchet}, {Carlberg}, {Nugent}, \& {Phillips}}]{hsiao07}
{Hsiao}, E.~Y., {Conley}, A., {Howell}, D.~A., {et~al.} 2007, \apj, 663, 1187

\bibitem[{{Jimenez} \& {Loeb}(2002)}]{jimenez02}
{Jimenez}, R., \& {Loeb}, A. 2002, \apj, 573, 37

\bibitem[{{Jones} {et~al.}(2013){Jones}, {Rodney}, {Riess}, {Mobasher},
  {Dahlen}, {McCully}, {Frederiksen}, {Casertano}, {Hjorth}, {Keeton},
  {Koekemoer}, {Strolger}, {Wiklind}, {Challis}, {Graur}, {Hayden}, {Patel},
  {Weiner}, {Filippenko}, {Garnavich}, {Jha}, {Kirshner}, {Ferguson}, {Grogin},
  \& {Kocevski}}]{jones13}
{Jones}, D.~O., {Rodney}, S.~A., {Riess}, A.~G., {et~al.} 2013, \apj, 768, 166

\bibitem[{{Kelly} {et~al.}(2010){Kelly}, {Hicken}, {Burke}, {Mandel}, \&
  {Kirshner}}]{kelly10}
{Kelly}, P.~L., {Hicken}, M., {Burke}, D.~L., {Mandel}, K.~S., \& {Kirshner},
  R.~P. 2010, \apj, 715, 743

\bibitem[{{Kelly} {et~al.}(2015){Kelly}, {Rodney}, {Treu}, {Foley}, {Brammer},
  {Schmidt}, {Zitrin}, {Sonnenfeld}, {Strolger}, {Graur}, {Filippenko}, {Jha},
  {Riess}, {Bradac}, {Weiner}, {Scolnic}, {Malkan}, {von der Linden}, {Trenti},
  {Hjorth}, {Gavazzi}, {Fontana}, {Merten}, {McCully}, {Jones}, {Postman},
  {Dressler}, {Patel}, {Cenko}, {Graham}, \& {Tucker}}]{kelly15}
{Kelly}, P.~L., {Rodney}, S.~A., {Treu}, T., {et~al.} 2015, Science, 347, 1123

\bibitem[{{Kelly} {et~al.}(2016{\natexlab{a}}){Kelly}, {Rodney}, {Treu},
  {Strolger}, {Foley}, {Jha}, {Selsing}, {Brammer}, {Brada{\v c}}, {Cenko},
  {Graur}, {Filippenko}, {Hjorth}, {McCully}, {Molino}, {Nonino}, {Riess},
  {Schmidt}, {Tucker}, {von der Linden}, {Weiner}, \& {Zitrin}}]{kelly16}
---. 2016{\natexlab{a}}, \apjl, 819, L8

\bibitem[{{Kelly} {et~al.}(2016{\natexlab{b}}){Kelly}, {Brammer}, {Selsing},
  {Foley}, {Hjorth}, {Rodney}, {Christensen}, {Strolger}, {Filippenko}, {Treu},
  {Steidel}, {Strom}, {Riess}, {Zitrin}, {Schmidt}, {Brada{\v c}}, {Jha},
  {Graham}, {McCully}, {Graur}, {Weiner}, {Silverman}, \& {Taddia}}]{kelly16b}
{Kelly}, P.~L., {Brammer}, G., {Selsing}, J., {et~al.} 2016{\natexlab{b}},
  \apj, 831, 205

\bibitem[{{Kessler} {et~al.}(2009){Kessler}, {Bernstein}, {Cinabro}, {Dilday},
  {Frieman}, {Jha}, {Kuhlmann}, {Miknaitis}, {Sako}, {Taylor}, \&
  {Vanderplas}}]{kessler09}
{Kessler}, R., {Bernstein}, J.~P., {Cinabro}, D., {et~al.} 2009, \pasp, 121,
  1028

\bibitem[{{Kessler} {et~al.}(2010){Kessler}, {Bassett}, {Belov}, {Bhatnagar},
  {Campbell}, {Conley}, {Frieman}, {Glazov}, {Gonz{\'a}lez-Gait{\'a}n},
  {Hlozek}, {Jha}, {Kuhlmann}, {Kunz}, {Lampeitl}, {Mahabal}, {Newling},
  {Nichol}, {Parkinson}, {Sajeeth Philip}, {Poznanski}, {Richards}, {Rodney},
  {Sako}, {Schneider}, {Smith}, {Stritzinger}, \& {Varughese}}]{kessler10}
{Kessler}, R., {Bassett}, B., {Belov}, P., {et~al.} 2010, \pasp, 122, 1415

\bibitem[{{Kim} {et~al.}(in prep.)}]{kim17}
{Kim}, S., {et~al.} in prep.

\bibitem[{{Kissler-Patig} {et~al.}(2008){Kissler-Patig}, {Pirard}, {Casali},
  {Moorwood}, {Ageorges}, {Alves de Oliveira}, {Baksai}, {Bedin}, {Bendek},
  {Biereichel}, {Delabre}, {Dorn}, {Esteves}, {Finger}, {Gojak}, {Huster},
  {Jung}, {Kiekebush}, {Klein}, {Koch}, {Lizon}, {Mehrgan}, {Petr-Gotzens},
  {Pritchard}, {Selman}, \& {Stegmeier}}]{kissler08}
{Kissler-Patig}, M., {Pirard}, J.-F., {Casali}, M., {et~al.} 2008, \aap, 491,
  941

\bibitem[{{Kriek} \& {Conroy}(2013)}]{kriek13}
{Kriek}, M., \& {Conroy}, C. 2013, \apjl, 775, L16

\bibitem[{{Kriek} {et~al.}(2009){Kriek}, {van Dokkum}, {Labb{\'e}}, {Franx},
  {Illingworth}, {Marchesini}, \& {Quadri}}]{kriek09}
{Kriek}, M., {van Dokkum}, P.~G., {Labb{\'e}}, I., {et~al.} 2009, \apj, 700,
  221

\bibitem[{{Kron}(1980)}]{kron80}
{Kron}, R.~G. 1980, \apjs, 43, 305

\bibitem[{{Levan} {et~al.}(2005){Levan}, {Nugent}, {Fruchter}, {Burud},
  {Branch}, {Rhoads}, {Castro-Tirado}, {Gorosabel}, {Castro Cer{\'o}n},
  {Thorsett}, {Kouveliotou}, {Golenetskii}, {Fynbo}, {Garnavich}, {Holland},
  {Hjorth}, {M{\o}ller}, {Pian}, {Tanvir}, {Ulanov}, {Wijers}, \&
  {Woosley}}]{levan05}
{Levan}, A., {Nugent}, P., {Fruchter}, A., {et~al.} 2005, \apj, 624, 880

\bibitem[{{Li} {et~al.}(2011){Li}, {Leaman}, {Chornock}, {Filippenko},
  {Poznanski}, {Ganeshalingam}, {Wang}, {Modjaz}, {Jha}, {Foley}, \&
  {Smith}}]{li11}
{Li}, W., {Leaman}, J., {Chornock}, R., {et~al.} 2011, \mnras, 412, 1441

\bibitem[{{Lochner} {et~al.}(2016){Lochner}, {McEwen}, {Peiris}, {Lahav}, \&
  {Winter}}]{lochner16}
{Lochner}, M., {McEwen}, J.~D., {Peiris}, H.~V., {Lahav}, O., \& {Winter},
  M.~K. 2016, ArXiv e-prints, arXiv:1603.00882

\bibitem[{{Maccoun} \& {Perlmutter}(2015)}]{maccoun15}
{Maccoun}, R., \& {Perlmutter}, S. 2015, \nat, 526, 187

\bibitem[{{Madau} \& {Dickinson}(2014)}]{madau14}
{Madau}, P., \& {Dickinson}, M. 2014, \araa, 52, 415

\bibitem[{{Mandel} {et~al.}(2016){Mandel}, {Scolnic}, {Shariff}, {Foley}, \&
  {Kirshner}}]{mandel16}
{Mandel}, K.~S., {Scolnic}, D., {Shariff}, H., {Foley}, R.~J., \& {Kirshner},
  R.~P. 2016, ArXiv e-prints, arXiv:1609.04470

\bibitem[{{Mannucci} {et~al.}(2006){Mannucci}, {Della Valle}, \&
  {Panagia}}]{mannucci06}
{Mannucci}, F., {Della Valle}, M., \& {Panagia}, N. 2006, \mnras, 370, 773

\bibitem[{{Maoz} \& {Mannucci}(2012)}]{maoz12}
{Maoz}, D., \& {Mannucci}, F. 2012, \pasa, 29, 447

\bibitem[{{Marchesini} {et~al.}(2007){Marchesini}, {van Dokkum}, {Quadri},
  {Rudnick}, {Franx}, {Lira}, {Wuyts}, {Gawiser}, {Christlein}, \&
  {Toft}}]{marchesini07}
{Marchesini}, D., {van Dokkum}, P., {Quadri}, R., {et~al.} 2007, \apj, 656, 42

\bibitem[{{Meyers} {et~al.}(2012){Meyers}, {Aldering}, {Barbary}, {Barrientos},
  {Brodwin}, {Dawson}, {Deustua}, {Doi}, {Eisenhardt}, {Faccioli}, {Fakhouri},
  {Fruchter}, {Gilbank}, {Gladders}, {Goldhaber}, {Gonzalez}, {Hattori},
  {Hsiao}, {Ihara}, {Kashikawa}, {Koester}, {Konishi}, {Lidman}, {Lubin},
  {Morokuma}, {Oda}, {Perlmutter}, {Postman}, {Ripoche}, {Rosati}, {Rubin},
  {Rykoff}, {Spadafora}, {Stanford}, {Suzuki}, {Takanashi}, {Tokita}, {Yasuda},
  \& {Supernova Cosmology Project}}]{meyers12}
{Meyers}, J., {Aldering}, G., {Barbary}, K., {et~al.} 2012, \apj, 750, 1

\bibitem[{{Moresco} {et~al.}(2012){Moresco}, {Cimatti}, {Jimenez}, {Pozzetti},
  {Zamorani}, {Bolzonella}, {Dunlop}, {Lamareille}, {Mignoli}, {Pearce},
  {Rosati}, {Stern}, {Verde}, {Zucca}, {Carollo}, {Contini}, {Kneib}, {Le
  F{\`e}vre}, {Lilly}, {Mainieri}, {Renzini}, {Scodeggio}, {Balestra}, {Gobat},
  {McLure}, {Bardelli}, {Bongiorno}, {Caputi}, {Cucciati}, {de la Torre}, {de
  Ravel}, {Franzetti}, {Garilli}, {Iovino}, {Kampczyk}, {Knobel}, {Kova{\v c}},
  {Le Borgne}, {Le Brun}, {Maier}, {Pell{\'o}}, {Peng}, {Perez-Montero},
  {Presotto}, {Silverman}, {Tanaka}, {Tasca}, {Tresse}, {Vergani}, {Almaini},
  {Barnes}, {Bordoloi}, {Bradshaw}, {Cappi}, {Chuter}, {Cirasuolo}, {Coppa},
  {Diener}, {Foucaud}, {Hartley}, {Kamionkowski}, {Koekemoer},
  {L{\'o}pez-Sanjuan}, {McCracken}, {Nair}, {Oesch}, {Stanford}, \&
  {Welikala}}]{moresco12}
{Moresco}, M., {Cimatti}, A., {Jimenez}, R., {et~al.} 2012, \jcap, 8, 006

\bibitem[{{Mosher} {et~al.}(2014){Mosher}, {Guy}, {Kessler}, {Astier},
  {Marriner}, {Betoule}, {Sako}, {El-Hage}, {Biswas}, {Pain}, {Kuhlmann},
  {Regnault}, {Frieman}, \& {Schneider}}]{mosher14}
{Mosher}, J., {Guy}, J., {Kessler}, R., {et~al.} 2014, \apj, 793, 16

\bibitem[{{Muzzin} {et~al.}(2012){Muzzin}, {Labb{\'e}}, {Franx}, {van Dokkum},
  {Holt}, {Szomoru}, {van de Sande}, {Brammer}, {Marchesini}, {Stefanon},
  {Buitrago}, {Caputi}, {Dunlop}, {Fynbo}, {Le F{\'e}vre}, {McCracken}, \&
  {Milvang-Jensen}}]{muzzin12}
{Muzzin}, A., {Labb{\'e}}, I., {Franx}, M., {et~al.} 2012, \apj, 761, 142

\bibitem[{{Navarro} {et~al.}(1997){Navarro}, {Frenk}, \& {White}}]{nfw97}
{Navarro}, J.~F., {Frenk}, C.~S., \& {White}, S.~D.~M. 1997, \apj, 490, 493

\bibitem[{{Newman} {et~al.}(2015){Newman}, {Belli}, \& {Ellis}}]{newman15}
{Newman}, A.~B., {Belli}, S., \& {Ellis}, R.~S. 2015, \apjl, 813, L7

\bibitem[{{Nordin} {et~al.}(2014){Nordin}, {Rubin}, {Richard}, {Rykoff},
  {Aldering}, {Amanullah}, {Atek}, {Barbary}, {Deustua}, {Fakhouri},
  {Fruchter}, {Goobar}, {Hook}, {Hsiao}, {Huang}, {Kneib}, {Lidman}, {Meyers},
  {Perlmutter}, {Saunders}, {Spadafora}, {Suzuki}, \& {Supernova Cosmology
  Project}}]{nordin14}
{Nordin}, J., {Rubin}, D., {Richard}, J., {et~al.} 2014, \mnras, 440, 2742

\bibitem[{{Nugent} {et~al.}(2002){Nugent}, {Kim}, \& {Perlmutter}}]{nugent02}
{Nugent}, P., {Kim}, A., \& {Perlmutter}, S. 2002, \pasp, 114, 803

\bibitem[{{Pandya} {et~al.}(2017){Pandya}, {Brennan}, {Somerville}, {Choi},
  {Barro}, {Wuyts}, {Taylor}, {Behroozi}, {Kirkpatrick}, {Faber}, {Primack},
  {Koo}, {McIntosh}, {Kocevski}, {Bell}, {Dekel}, {Fang}, {Ferguson}, {Grogin},
  {Koekemoer}, {Lu}, {Mantha}, {Mobasher}, {Newman}, {Pacifici}, {Papovich},
  {van der Wel}, \& {Yesuf}}]{pandya17}
{Pandya}, V., {Brennan}, R., {Somerville}, R.~S., {et~al.} 2017, \mnras, 472,
  2054

\bibitem[{{Patel} {et~al.}(2014){Patel}, {McCully}, {Jha}, {Rodney}, {Jones},
  {Graur}, {Merten}, {Zitrin}, {Riess}, {Matheson}, {Sako}, {Holoien},
  {Postman}, {Coe}, {Bartelmann}, {Balestra}, {Ben{\'{\i}}tez}, {Bouwens},
  {Bradley}, {Broadhurst}, {Cenko}, {Donahue}, {Filippenko}, {Ford},
  {Garnavich}, {Grillo}, {Infante}, {Jouvel}, {Kelson}, {Koekemoer}, {Lahav},
  {Lemze}, {Maoz}, {Medezinski}, {Melchior}, {Meneghetti}, {Molino},
  {Moustakas}, {Moustakas}, {Nonino}, {Rosati}, {Seitz}, {Strolger}, {Umetsu},
  \& {Zheng}}]{patel14}
{Patel}, B., {McCully}, C., {Jha}, S.~W., {et~al.} 2014, \apj, 786, 9

\bibitem[{{Petrushevska} {et~al.}(2016){Petrushevska}, {Amanullah}, {Goobar},
  {Fabbro}, {Johansson}, {Kjellsson}, {Lidman}, {Paech}, {Richard}, {Dahle},
  {Ferretti}, {Kneib}, {Limousin}, {Nordin}, \& {Stanishev}}]{petrushevska16}
{Petrushevska}, T., {Amanullah}, R., {Goobar}, A., {et~al.} 2016, \aap, 594,
  A54

\bibitem[{{Planck Collaboration} {et~al.}(2015){Planck Collaboration}, {Ade},
  {Aghanim}, {Arnaud}, {Ashdown}, {Aumont}, {Baccigalupi}, {Banday},
  {Barreiro}, {Bartlett}, \& et~al.}]{planck15}
{Planck Collaboration}, {Ade}, P.~A.~R., {Aghanim}, N., {et~al.} 2015, ArXiv
  e-prints, arXiv:1502.01589

\bibitem[{{Quimby} {et~al.}(2014){Quimby}, {Oguri}, {More}, {More}, {Moriya},
  {Werner}, {Tanaka}, {Folatelli}, {Bersten}, {Maeda}, \& {Nomoto}}]{quimby14}
{Quimby}, R.~M., {Oguri}, M., {More}, A., {et~al.} 2014, Science, 344, 396

\bibitem[{{Rafelski} {et~al.}(2015){Rafelski}, {Teplitz}, {Gardner}, {Coe},
  {Bond}, {Koekemoer}, {Grogin}, {Kurczynski}, {McGrath}, {Bourque}, {Atek},
  {Brown}, {Colbert}, {Codoreanu}, {Ferguson}, {Finkelstein}, {Gawiser},
  {Giavalisco}, {Gronwall}, {Hanish}, {Lee}, {Mehta}, {de Mello},
  {Ravindranath}, {Ryan}, {Scarlata}, {Siana}, {Soto}, \& {Voyer}}]{rafelski15}
{Rafelski}, M., {Teplitz}, H.~I., {Gardner}, J.~P., {et~al.} 2015, \aj, 150, 31

\bibitem[{{Richardson} {et~al.}(2014){Richardson}, {Jenkins}, {Wright}, \&
  {Maddox}}]{richardson14}
{Richardson}, D., {Jenkins}, III, R.~L., {Wright}, J., \& {Maddox}, L. 2014,
  \aj, 147, 118

\bibitem[{{Rigault} {et~al.}(2013){Rigault}, {Copin}, {Aldering}, {Antilogus},
  {Aragon}, {Bailey}, {Baltay}, {Bongard}, {Buton}, {Canto}, {Cellier-Holzem},
  {Childress}, {Chotard}, {Fakhouri}, {Feindt}, {Fleury}, {Gangler},
  {Greskovic}, {Guy}, {Kim}, {Kowalski}, {Lombardo}, {Nordin}, {Nugent},
  {Pain}, {P{\'e}contal}, {Pereira}, {Perlmutter}, {Rabinowitz}, {Runge},
  {Saunders}, {Scalzo}, {Smadja}, {Tao}, {Thomas}, \& {Weaver}}]{rigault13}
{Rigault}, M., {Copin}, Y., {Aldering}, G., {et~al.} 2013, \aap, 560, A66

\bibitem[{{Rodney} \& {Tonry}(2009)}]{rodney09}
{Rodney}, S.~A., \& {Tonry}, J.~L. 2009, \apj, 707, 1064

\bibitem[{{Rodney} {et~al.}(2012){Rodney}, {Riess}, {Dahlen}, {Strolger},
  {Ferguson}, {Hjorth}, {Frederiksen}, {Weiner}, {Mobasher}, {Casertano},
  {Jones}, {Challis}, {Faber}, {Filippenko}, {Garnavich}, {Graur}, {Grogin},
  {Hayden}, {Jha}, {Kirshner}, {Kocevski}, {Koekemoer}, {McCully}, {Patel},
  {Rajan}, \& {Scarlata}}]{rodney12}
{Rodney}, S.~A., {Riess}, A.~G., {Dahlen}, T., {et~al.} 2012, \apj, 746, 5

\bibitem[{{Rodney} {et~al.}(2014){Rodney}, {Riess}, {Strolger}, {Dahlen},
  {Graur}, {Casertano}, {Dickinson}, {Ferguson}, {Garnavich}, {Hayden}, {Jha},
  {Jones}, {Kirshner}, {Koekemoer}, {McCully}, {Mobasher}, {Patel}, {Weiner},
  {Cenko}, {Clubb}, {Cooper}, {Filippenko}, {Frederiksen}, {Hjorth},
  {Leibundgut}, {Matheson}, {Nayyeri}, {Penner}, {Trump}, {Silverman}, {U},
  {Azalee Bostroem}, {Challis}, {Rajan}, {Wolff}, {Faber}, {Grogin}, \&
  {Kocevski}}]{rodney14}
{Rodney}, S.~A., {Riess}, A.~G., {Strolger}, L.-G., {et~al.} 2014, \aj, 148, 13

\bibitem[{{Rodney} {et~al.}(2015{\natexlab{a}}){Rodney}, {Patel}, {Scolnic},
  {Foley}, {Molino}, {Brammer}, {Jauzac}, {Brada{\v c}}, {Broadhurst}, {Coe},
  {Diego}, {Graur}, {Hjorth}, {Hoag}, {Jha}, {Johnson}, {Kelly}, {Lam},
  {McCully}, {Medezinski}, {Meneghetti}, {Merten}, {Richard}, {Riess},
  {Sharon}, {Strolger}, {Treu}, {Wang}, {Williams}, \& {Zitrin}}]{rodney15}
{Rodney}, S.~A., {Patel}, B., {Scolnic}, D., {et~al.} 2015{\natexlab{a}}, \apj,
  811, 70

\bibitem[{{Rodney} {et~al.}(2015{\natexlab{b}}){Rodney}, {Riess}, {Scolnic},
  {Jones}, {Hemmati}, {Molino}, {McCully}, {Mobasher}, {Strolger}, {Graur},
  {Hayden}, \& {Casertano}}]{rodney15b}
{Rodney}, S.~A., {Riess}, A.~G., {Scolnic}, D.~M., {et~al.} 2015{\natexlab{b}},
  \aj, 150, 156

\bibitem[{{Rubin} {et~al.}(2013){Rubin}, {Knop}, {Rykoff}, {Aldering},
  {Amanullah}, {Barbary}, {Burns}, {Conley}, {Connolly}, {Deustua}, {Fadeyev},
  {Fakhouri}, {Fruchter}, {Gibbons}, {Goldhaber}, {Goobar}, {Hsiao}, {Huang},
  {Kowalski}, {Lidman}, {Meyers}, {Nordin}, {Perlmutter}, {Saunders},
  {Spadafora}, {Stanishev}, {Suzuki}, {Wang}, \& {Supernova Cosmology
  Project}}]{rubin13}
{Rubin}, D., {Knop}, R.~A., {Rykoff}, E., {et~al.} 2013, \apj, 763, 35

\bibitem[{{Rubin} {et~al.}(2015){Rubin}, {Aldering}, {Barbary}, {Boone},
  {Chappell}, {Currie}, {Deustua}, {Fagrelius}, {Fruchter}, {Hayden}, {Lidman},
  {Nordin}, {Perlmutter}, {Saunders}, {Sofiatti}, \& {Supernova Cosmology
  Project}}]{rubin15}
{Rubin}, D., {Aldering}, G., {Barbary}, K., {et~al.} 2015, \apj, 813, 137

\bibitem[{{Rubin} {et~al.}(in prep.)}]{rubin16b}
{Rubin}, D., {et~al.} in prep.

\bibitem[{{Sako} {et~al.}(2011){Sako}, {Bassett}, {Connolly}, {Dilday},
  {Cambell}, {Frieman}, {Gladney}, {Kessler}, {Lampeitl}, {Marriner}, {Miquel},
  {Nichol}, {Schneider}, {Smith}, \& {Sollerman}}]{sako11}
{Sako}, M., {Bassett}, B., {Connolly}, B., {et~al.} 2011, \apj, 738, 162

\bibitem[{{Salpeter}(1955)}]{salpeter55}
{Salpeter}, E.~E. 1955, \apj, 121, 161

\bibitem[{{Scalzo} {et~al.}(2012){Scalzo}, {Aldering}, {Antilogus}, {Aragon},
  {Bailey}, {Baltay}, {Bongard}, {Buton}, {Canto}, {Cellier-Holzem},
  {Childress}, {Chotard}, {Copin}, {Fakhouri}, {Gangler}, {Guy}, {Hsiao},
  {Kerschhaggl}, {Kowalski}, {Nugent}, {Paech}, {Pain}, {Pecontal}, {Pereira},
  {Perlmutter}, {Rabinowitz}, {Rigault}, {Runge}, {Smadja}, {Tao}, {Thomas},
  {Weaver}, {Wu}, \& {Nearby Supernova Factory}}]{scalzo12}
{Scalzo}, R., {Aldering}, G., {Antilogus}, P., {et~al.} 2012, \apj, 757, 12

\bibitem[{{Scannapieco} \& {Bildsten}(2005)}]{scannapieco05}
{Scannapieco}, E., \& {Bildsten}, L. 2005, \apjl, 629, L85

\bibitem[{{Schlafly} {et~al.}(2016){Schlafly}, {Meisner}, {Stutz},
  {Kainulainen}, {Peek}, {Tchernyshyov}, {Rix}, {Finkbeiner}, {Covey}, {Green},
  {Bell}, {Burgett}, {Chambers}, {Draper}, {Flewelling}, {Hodapp}, {Kaiser},
  {Magnier}, {Martin}, {Metcalfe}, {Wainscoat}, \& {Waters}}]{schlafly16}
{Schlafly}, E.~F., {Meisner}, A.~M., {Stutz}, A.~M., {et~al.} 2016, \apj, 821,
  78

\bibitem[{{Scolnic} \& {Kessler}(2016)}]{scolnic16}
{Scolnic}, D., \& {Kessler}, R. 2016, \apjl, 822, L35

\bibitem[{{Scolnic} {et~al.}(2014){Scolnic}, {Riess}, {Foley}, {Rest},
  {Rodney}, {Brout}, \& {Jones}}]{scolnic14}
{Scolnic}, D.~M., {Riess}, A.~G., {Foley}, R.~J., {et~al.} 2014, \apj, 780, 37

\bibitem[{{S{\'e}rsic}(1963)}]{sersic63}
{S{\'e}rsic}, J.~L. 1963, Boletin de la Asociacion Argentina de Astronomia La
  Plata Argentina, 6, 41

\bibitem[{{Silverman} {et~al.}(2012){Silverman}, {Kong}, \&
  {Filippenko}}]{silverman12}
{Silverman}, J.~M., {Kong}, J.~J., \& {Filippenko}, A.~V. 2012, \mnras, 425,
  1819

\bibitem[{{Skrutskie} {et~al.}(2006){Skrutskie}, {Cutri}, {Stiening},
  {Weinberg}, {Schneider}, {Carpenter}, {Beichman}, {Capps}, {Chester},
  {Elias}, {Huchra}, {Liebert}, {Lonsdale}, {Monet}, {Price}, {Seitzer},
  {Jarrett}, {Kirkpatrick}, {Gizis}, {Howard}, {Evans}, {Fowler}, {Fullmer},
  {Hurt}, {Light}, {Kopan}, {Marsh}, {McCallon}, {Tam}, {Van Dyk}, \&
  {Wheelock}}]{skrutskie06}
{Skrutskie}, M.~F., {Cutri}, R.~M., {Stiening}, R., {et~al.} 2006, \aj, 131,
  1163

\bibitem[{{Stanford} {et~al.}(2014){Stanford}, {Gonzalez}, {Brodwin},
  {Gettings}, {Eisenhardt}, {Stern}, \& {Wylezalek}}]{stanford14}
{Stanford}, S.~A., {Gonzalez}, A.~H., {Brodwin}, M., {et~al.} 2014, \apjs, 213,
  25

\bibitem[{{Stanishev} {et~al.}(2009){Stanishev}, {Goobar}, {Paech},
  {Amanullah}, {Dahl{\'e}n}, {J{\"o}nsson}, {Kneib}, {Lidman}, {Limousin},
  {M{\"o}rtsell}, {Nobili}, {Richard}, {Riehm}, \& {von Strauss}}]{Stanishev09}
{Stanishev}, V., {Goobar}, A., {Paech}, K., {et~al.} 2009, \aap, 507, 61

\bibitem[{{Stern} {et~al.}(2010){Stern}, {Jimenez}, {Verde}, {Kamionkowski}, \&
  {Stanford}}]{stern10}
{Stern}, D., {Jimenez}, R., {Verde}, L., {Kamionkowski}, M., \& {Stanford},
  S.~A. 2010, \jcap, 2, 008

\bibitem[{{Stern} {et~al.}(2004){Stern}, {van Dokkum}, {Nugent}, {Sand},
  {Ellis}, {Sullivan}, {Bloom}, {Frail}, {Kneib}, {Koopmans}, \&
  {Treu}}]{stern04}
{Stern}, D., {van Dokkum}, P.~G., {Nugent}, P., {et~al.} 2004, \apj, 612, 690

\bibitem[{{Strolger} {et~al.}(2015){Strolger}, {Dahlen}, {Rodney}, {Graur},
  {Riess}, {McCully}, {Ravindranath}, {Mobasher}, \& {Shahady}}]{strolger15}
{Strolger}, L.-G., {Dahlen}, T., {Rodney}, S.~A., {et~al.} 2015, \apj, 813, 93

\bibitem[{{Sullivan} {et~al.}(2006){Sullivan}, {Le Borgne}, {Pritchet},
  {Hodsman}, {Neill}, {Howell}, {Carlberg}, {Astier}, {Aubourg}, {Balam},
  {Basa}, {Conley}, {Fabbro}, {Fouchez}, {Guy}, {Hook}, {Pain},
  {Palanque-Delabrouille}, {Perrett}, {Regnault}, {Rich}, {Taillet}, {Baumont},
  {Bronder}, {Ellis}, {Filiol}, {Lusset}, {Perlmutter}, {Ripoche}, \&
  {Tao}}]{sullivan06}
{Sullivan}, M., {Le Borgne}, D., {Pritchet}, C.~J., {et~al.} 2006, \apj, 648,
  868

\bibitem[{{Sullivan} {et~al.}(2010){Sullivan}, {Conley}, {Howell}, {Neill},
  {Astier}, {Balland}, {Basa}, {Carlberg}, {Fouchez}, {Guy}, {Hardin}, {Hook},
  {Pain}, {Palanque-Delabrouille}, {Perrett}, {Pritchet}, {Regnault}, {Rich},
  {Ruhlmann-Kleider}, {Baumont}, {Hsiao}, {Kronborg}, {Lidman}, {Perlmutter},
  \& {Walker}}]{sullivan10}
{Sullivan}, M., {Conley}, A., {Howell}, D.~A., {et~al.} 2010, \mnras, 406, 782

\bibitem[{{Suzuki} {et~al.}(2012){Suzuki}, {Rubin}, {Lidman}, {Aldering},
  {Amanullah}, {Barbary}, {Barrientos}, {Botyanszki}, {Brodwin}, {Connolly},
  {Dawson}, {Dey}, {Doi}, {Donahue}, {Deustua}, {Eisenhardt}, {Ellingson},
  {Faccioli}, {Fadeyev}, {Fakhouri}, {Fruchter}, {Gilbank}, {Gladders},
  {Goldhaber}, {Gonzalez}, {Goobar}, {Gude}, {Hattori}, {Hoekstra}, {Hsiao},
  {Huang}, {Ihara}, {Jee}, {Johnston}, {Kashikawa}, {Koester}, {Konishi},
  {Kowalski}, {Linder}, {Lubin}, {Melbourne}, {Meyers}, {Morokuma}, {Munshi},
  {Mullis}, {Oda}, {Panagia}, {Perlmutter}, {Postman}, {Pritchard}, {Rhodes},
  {Ripoche}, {Rosati}, {Schlegel}, {Spadafora}, {Stanford}, {Stanishev},
  {Stern}, {Strovink}, {Takanashi}, {Tokita}, {Wagner}, {Wang}, {Yasuda},
  {Yee}, \& {Supernova Cosmology Project}}]{suzuki12}
{Suzuki}, N., {Rubin}, D., {Lidman}, C., {et~al.} 2012, \apj, 746, 85

\bibitem[{{Tody}(1993)}]{1993ASPC...52..173T}
{Tody}, D. 1993, in Astronomical Society of the Pacific Conference Series,
  Vol.~52, Astronomical Data Analysis Software and Systems II, ed. R.~J.
  {Hanisch}, R.~J.~V. {Brissenden}, \& J.~{Barnes}, 173

\bibitem[{{Tripp}(1998)}]{tripp98}
{Tripp}, R. 1998, \aap, 331, 815

\bibitem[{{van der Wel} {et~al.}(2014){van der Wel}, {Franx}, {van Dokkum},
  {Skelton}, {Momcheva}, {Whitaker}, {Brammer}, {Bell}, {Rix}, {Wuyts},
  {Ferguson}, {Holden}, {Barro}, {Koekemoer}, {Chang}, {McGrath},
  {H{\"a}ussler}, {Dekel}, {Behroozi}, {Fumagalli}, {Leja}, {Lundgren},
  {Maseda}, {Nelson}, {Wake}, {Patel}, {Labb{\'e}}, {Faber}, {Grogin}, \&
  {Kocevski}}]{vanderwel14}
{van der Wel}, A., {Franx}, M., {van Dokkum}, P.~G., {et~al.} 2014, \apj, 788,
  28

\bibitem[{{van Dokkum} {et~al.}(2015){van Dokkum}, {Nelson}, {Franx}, {Oesch},
  {Momcheva}, {Brammer}, {F{\"o}rster Schreiber}, {Skelton}, {Whitaker}, {van
  der Wel}, {Bezanson}, {Fumagalli}, {Illingworth}, {Kriek}, {Leja}, \&
  {Wuyts}}]{vandokkum15}
{van Dokkum}, P.~G., {Nelson}, E.~J., {Franx}, M., {et~al.} 2015, \apj, 813, 23

\bibitem[{{Vernet} {et~al.}(2011){Vernet}, {Dekker}, {D'Odorico}, {Kaper},
  {Kjaergaard}, {Hammer}, {Randich}, {Zerbi}, {Groot}, {Hjorth}, {Guinouard},
  {Navarro}, {Adolfse}, {Albers}, {Amans}, {Andersen}, {Andersen}, {Binetruy},
  {Bristow}, {Castillo}, {Chemla}, {Christensen}, {Conconi}, {Conzelmann},
  {Dam}, {de Caprio}, {de Ugarte Postigo}, {Delabre}, {di Marcantonio},
  {Downing}, {Elswijk}, {Finger}, {Fischer}, {Flores}, {Fran{\c c}ois},
  {Goldoni}, {Guglielmi}, {Haigron}, {Hanenburg}, {Hendriks}, {Horrobin},
  {Horville}, {Jessen}, {Kerber}, {Kern}, {Kiekebusch}, {Kleszcz}, {Klougart},
  {Kragt}, {Larsen}, {Lizon}, {Lucuix}, {Mainieri}, {Manuputy}, {Martayan},
  {Mason}, {Mazzoleni}, {Michaelsen}, {Modigliani}, {Moehler}, {M{\o}ller},
  {Norup S{\o}rensen}, {N{\o}rregaard}, {P{\'e}roux}, {Patat}, {Pena}, {Pragt},
  {Reinero}, {Rigal}, {Riva}, {Roelfsema}, {Royer}, {Sacco}, {Santin},
  {Schoenmaker}, {Spano}, {Sweers}, {Ter Horst}, {Tintori}, {Tromp}, {van
  Dael}, {van der Vliet}, {Venema}, {Vidali}, {Vinther}, {Vola}, {Winters},
  {Wistisen}, {Wulterkens}, \& {Zacchei}}]{vernet11}
{Vernet}, J., {Dekker}, H., {D'Odorico}, S., {et~al.} 2011, \aap, 536, A105

\bibitem[{{Whitaker} {et~al.}(2013){Whitaker}, {van Dokkum}, {Brammer},
  {Momcheva}, {Skelton}, {Franx}, {Kriek}, {Labb{\'e}}, {Fumagalli},
  {Lundgren}, {Nelson}, {Patel}, \& {Rix}}]{whitaker13}
{Whitaker}, K.~E., {van Dokkum}, P.~G., {Brammer}, G., {et~al.} 2013, \apjl,
  770, L39

\bibitem[{{Williams} {et~al.}(2009){Williams}, {Quadri}, {Franx}, {van Dokkum},
  \& {Labb{\'e}}}]{williams09}
{Williams}, R.~J., {Quadri}, R.~F., {Franx}, M., {van Dokkum}, P., \&
  {Labb{\'e}}, I. 2009, \apj, 691, 1879

\bibitem[{{Wright} {et~al.}(2010){Wright}, {Eisenhardt}, {Mainzer}, {Ressler},
  {Cutri}, {Jarrett}, {Kirkpatrick}, {Padgett}, {McMillan}, {Skrutskie},
  {Stanford}, {Cohen}, {Walker}, {Mather}, {Leisawitz}, {Gautier}, {McLean},
  {Benford}, {Lonsdale}, {Blain}, {Mendez}, {Irace}, {Duval}, {Liu}, {Royer},
  {Heinrichsen}, {Howard}, {Shannon}, {Kendall}, {Walsh}, {Larsen}, {Cardon},
  {Schick}, {Schwalm}, {Abid}, {Fabinsky}, {Naes}, \& {Tsai}}]{wright10}
{Wright}, E.~L., {Eisenhardt}, P.~R.~M., {Mainzer}, A.~K., {et~al.} 2010, \aj,
  140, 1868

\end{thebibliography}
\end{document}